\documentclass[fleqn,usenatbib]{mnras}

\usepackage{amsmath}	
\usepackage{txfonts,times}

\usepackage[T1]{fontenc}
\usepackage{ae,aecompl}

\usepackage{graphicx}	

\title[The optical transmission spectrum of WASP-31b]{VLT/FORS2 comparative transmission spectroscopy II: confirmation of a cloud-deck and Rayleigh scattering in WASP-31b, but no potassium?}

\author[N. P. Gibson et al.]{
Neale P. Gibson$^{1}$\thanks{E-mail: n.gibson@qub.ac.uk},
Nikolay Nikolov$^{2}$,
David K. Sing$^{2}$,
Joanna K. Barstow$^{3}$,  \newauthor
Thomas M. Evans$^{2}$,
Tiffany Kataria$^{4}$ and
Paul A. Wilson$^{5}$
\smallskip
\\
$^{1}$Astrophysics Research Centre, School of Mathematics and Physics, Queens University Belfast, Belfast BT7 1NN, UK\\
$^{2}$Physics and Astronomy, University of Exeter, Exeter EX4 4QL,  UK\\
$^{3}$Physics and Astronomy, University College London, London NW1 2PS, UK\\
$^{4}$NASA Jet Propulsion Laboratory, 4800 Oak Grove Dr, Pasadena, CA 91109, USA\\
$^{5}$Institut d'Astrophysique de Paris, 98$^{\rm{bis}}$ Boulevard Arago, F-75014 Paris, France\\
}

\date{Accepted XXX. Received YYY; in original form ZZZ}

\pubyear{2016}

\begin{document}
\label{firstpage}
\pagerange{\pageref{firstpage}--\pageref{lastpage}}
\maketitle

\begin{abstract}
We present transmission spectroscopy of the hot-Jupiter WASP-31b using FORS2 on the {VLT} during two primary transits. The observations cover a wavelength range of $\approx$400--840\,nm. The light curves are corrupted by significant systematics, but these were to first order invariant with wavelength and could be removed using a common-mode correction derived from the white light curves. We reach a precision in the transit depth of $\approx$140 ppm in 15\,nm bins, although the precision varies significantly over the wavelength range. Our FORS2 observations confirm the cloud-deck previously inferred using {\it HST}/STIS. We also re-analyse the {\it HST}/STIS data using a Gaussian process model, finding excellent agreement with earlier measurements. We reproduce the Rayleigh scattering signature at short wavelengths ($\lesssim$\,$5300$\,\AA) and the cloud-deck at longer wavelengths. However, our FORS2 observations appear to rule out the large potassium feature previously detected using STIS, yet it is recovered from the {\it HST}/STIS data, although with reduced amplitude and significance ($\approx$\,$2.5\sigma$). The discrepancy between our results and the earlier STIS detection of potassium ($\approx$\,$4.3\sigma$) is either a result of telluric contamination of the ground-based observations, or an underestimate of the uncertainties for narrow-band features in {\it HST}/STIS when using linear basis models to account for the systematics. Our results further demonstrate the use of ground-based multi-object spectrographs for the study of exoplanet atmospheres, and highlight the need for caution in our interpretation of narrow-band features in low-resolution spectra of hot-Jupiters.

\end{abstract}




\section{Introduction}

Transit and radial velocity surveys have made remarkable progress in understanding the population of extrasolar planets in our Galaxy. Obtaining spectroscopy of them is the next step in understanding the chemistry and physical processes in their atmospheres. Transiting planets enable such measurements by {\it temporally} resolving planets from their host stars, rather than spatially separating light from the planet and star. One such technique, transmission spectroscopy, measures the apparent radius of a planet during primary transit as a function of wavelength. This is the altitude at which the planet becomes opaque to starlight, which depends on the opacity, and therefore the composition and physical structure of the planet's atmosphere \citep{Seager_2000,Brown_2001}.

Space-based observations have led the way in our understanding of exoplanet atmospheres \citep[e.g.][]{Charbonneau_2002,Pont_2008,Huitson_2012,Berta_2012,Pont_2013,Kreidberg_2014,Nikolov_2015,Sing_2016}, however ground-based observations are rapidly increasing in importance, with the adoption of multi-object spectrographs (MOS) to perform differential spectro-photometry \citep[e.g.][]{Bean_2010,Bean_2011,Crossfield_2013,Gibson_2013a,Gibson_2013b,Jordan_2013,Kirk_2016,Lendl_2016,Mallonn_2016,Stevenson_2014b}. The FOcal Reducer and low dispersion Spectrograph \citep[FORS2;][]{Appenzeller_1998} on the European Southern Observatory's (ESO) Very Large Telescope (VLT) was the first to demonstrate this technique successfully \citep{Bean_2010}, but its impact has been limited by significant systematic effects related to the non-uniformity of the Linear Atmospheric Dispersion Corrector (LADC). This component has been recently upgraded \citep{Boffin_2015,Boffin_2016}, and \citet{Sedaghati_2015} has since demonstrated the improved performance of the instrument for exoplanet spectroscopy.

Here we report on the use of FORS2 to observe the transmission spectrum of the hot-Jupiter, WASP-31b. This is part of a small survey to re-observe targets that we have already observed using the {\it Hubble Space Telescope (HST)} \citep{Sing_2016}, and our first results for WASP-39b are already reported in \citet{Nikolov_2016}. We selected targets for which their spectra already show evidence for alkali absorption and scattering by aerosols that should be easily detectable from the ground. Our survey has two general aims: to test the performance of FORS2 by observing objects with known spectroscopic features; and to confirm the signals detected with {\it HST}. This is particularly important, as the field of exoplanet spectroscopy has been limited by our ability to model instrumental time-series systematics \citep[e.g.][]{Gibson_2011,Gibson_2012}, and observing planets with multiple telescopes gives us the opportunity to verify and refine our current results and methodology.

WASP-31b is an inflated hot-Jupiter discovered by \citet{Anderson_2011}, with a mass and radius of $\approx0.48\,M_{\rm J}$ and $1.55\,R_{\rm J}$, respectively. It orbits a late F-type dwarf ($V=11.7$) with a period of 3.4\,days. The optical and near-infrared transmission spectra have already been observed by \citet{Sing_2015} using the Space Telescope Imaging Spectrograph (STIS) and Wide Field Camera 3 (WFC3), revealing strong Rayleigh- and Mie-scattering by aerosols, and a strong K feature (but no Na), that perhaps indicates a sub-stellar Na/K abundance ratio, and has interesting implications for the formation and/or evolution of the planet. These features make this an excellent target for our FORS2 observations, which should be able to recover and confirm these features. This paper is structured as follows; in Sect.~\ref{sect:observations} we describe the observations and data reduction and in Sect.~\ref{sect:analysis} we present our light curve analysis and extraction of the transmission spectra. Finally in Sects.~\ref{sect:discussion} and \ref{sect:conclusion} we present our results and conclusions.

\section{FORS2 Observations}
\label{sect:observations}
Two transits of WASP-31b were observed using the 8.2-m `Antu' telescope (Unit Telescope 1) of the VLT with FORS2: a general purpose imager, (multi-object) spectrograph and polarimeter. Observations were taken on the nights of 2016 February 15 and 2016 March 3, as part of program 096.C-0765 (PI: Nikolov), and followed a similar observing strategy to that described in \citet{Nikolov_2016}.  The first transit was observed using the grism GRIS600B (hereafter 600B). Science exposures covered 5.2 hours, with 266 exposures of 40s (except the first 2). The second transit was observed using the grism GRIS600RI (hereafter 600RI) in combination with the GG435 order blocking filter, and science exposures lasted 5 hours, with 325 exposures typically ranging from 25--30s, and were adjusted through the night due to varying conditions.

\begin{figure}
\centering
\includegraphics[width=76mm]{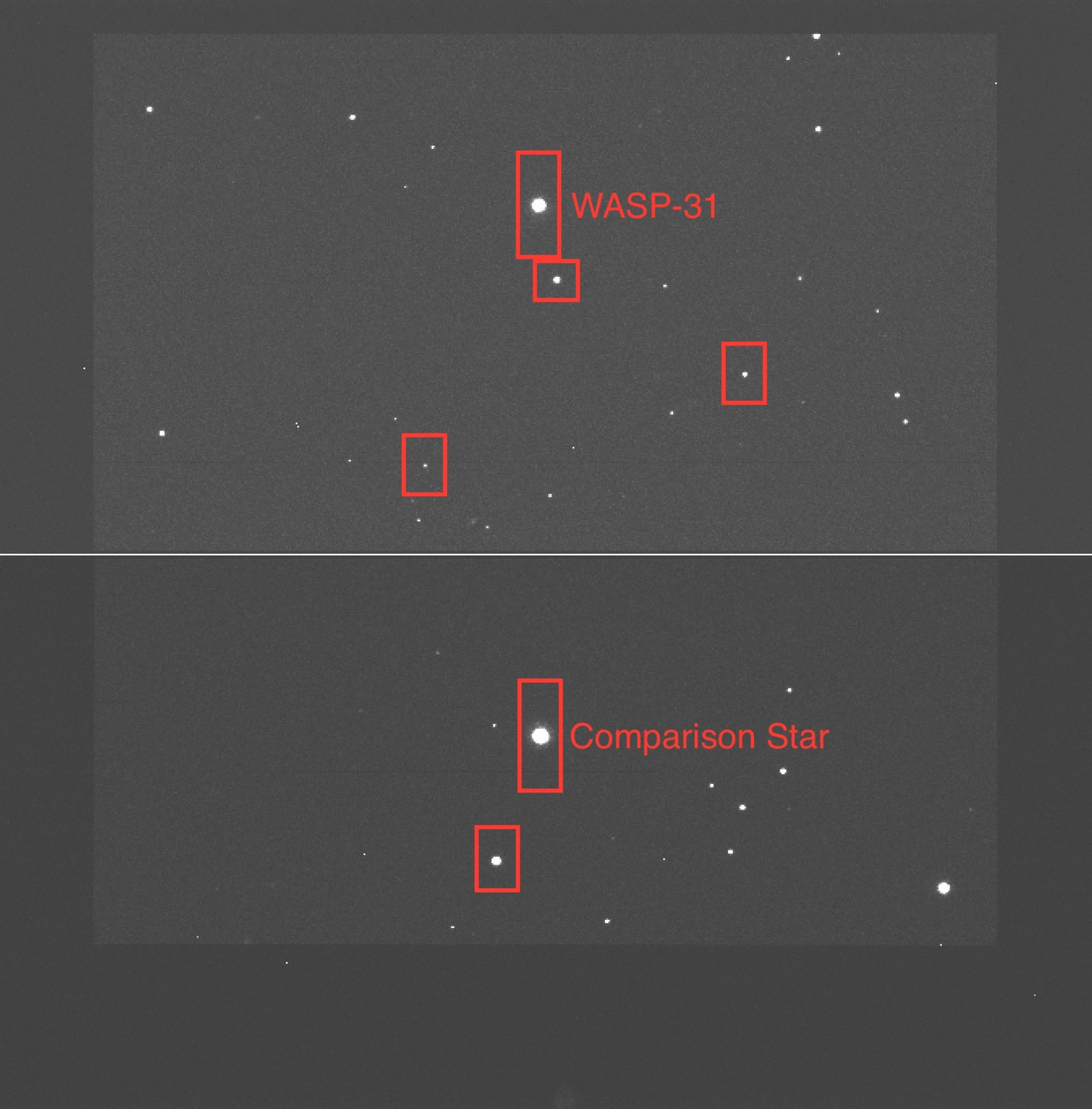}
\caption{FORS2 acquisition image, showing the detector gap, the approximate positions and sizes of the science slits, and the position of the target and main acquisition star. The dispersion axis is horizontal.}
\label{fig:fov}
\end{figure}

FORS2 has an imaging field-of-view of 6.8$\times$6.8 arcminutes squared, and consists of two 2k $\times$ 4k CCDs arranged with a small (few arcsecond) gap in the cross-dispersion axis. We observed the target (V$\sim$11.7) and 5 comparison stars simultaneously in multi-object (MXU) mode, using a custom mask designed from FORS2 pre-imaging to accurately place the slits. Observations were taken in $2\times2$ binning mode, giving a pixel scale of 0.25$^{\prime\prime}$/pixel. For our analysis, only the brightest comparison star was used (V$\sim$11.4), as the other stars were significantly fainter and did not affect the final light curves. An acquisition image is shown in Fig.~\ref{fig:fov}, showing the arrangement of the CCDs, the target and comparison star, and the approximate positions and shape of the slits. The width of all slits was 22$^{\prime\prime}$, and the lengths 50$^{\prime\prime}$ and 65$^{\prime\prime}$ for the target and comparison star, respectively. This resulted in seeing-limited resolution of $R\approx 450-1050$ for the first night (FWHM varied from 3--7 pixels, at $\lambda\approx4700\AA$), and $R\approx1400-2100$ for the second night (FWHM$\approx$2--3 pixels, at $\lambda\approx6800\AA$). We also constructed a calibration mask matching the science mask, but with 1$^{\prime\prime}$ slit widths. This was used to obtain flat fields that would more closely match the stellar PSF than the wide slit, and arcs with narrower features for more precise wavelength calibration.

Data were bias subtracted and flat fielded using the FORS2 pipeline; however, these procedures made little difference to the final results, and we decided to extract spectra directly from the raw frames. This is not surprising, given that stable spatial variations in the detector sensitivity will cancel out in the differential photometry, and there were no high spatial frequency variations in the flat field structure. Furthermore, it is impossible to properly flat-field the data using a single flat field, as this cannot account for the difference in the slit transmission (for the sky background) and the stellar PSF, nor changes in the shape and position of the stellar PSFs that arise due to pointing errors and seeing variation. In the end the FORS2 pipeline was only used for the wavelength calibration, which used arc lamp exposures taken with the calibration mask.

\begin{figure*}
\centering
\includegraphics[width=170mm]{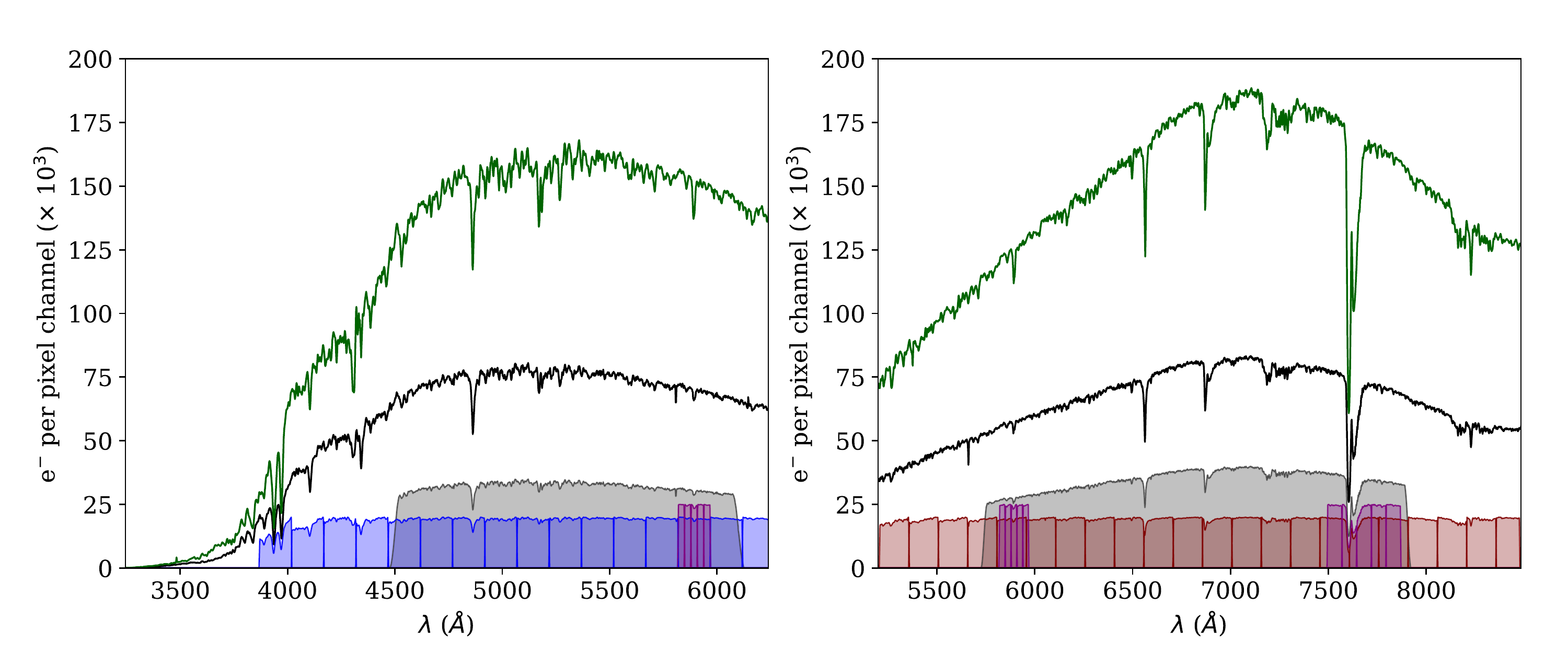}
\caption{Example spectra of WASP-31 (black) and the comparison star (green). The left and right panels show the 600B and the 600RI data, respectively. The lower panels show the effective spectral bins (i.e. spectral window times the target spectrum, arbitrary scaling) used to construct the white light curves (grey), spectral channels (blue and red), and the high-resolution channels around narrow Na and K features (purple, see text for details).}
\label{fig:spectra}
\end{figure*}

The spectra for the target and comparison stars were extracted using a custom pipeline written using {\sc IRAF}\footnote{{\sc IRAF} is distributed by the National Optical Astronomy Observatory, which is operated by the Association of Universities for Research in Astronomy (AURA) under cooperative agreement with the National Science Foundation}/{\sc PyRAF}\footnote{{\sc PyRAF} is a product of the Space Telescope Science Institute, which is operated by AURA for NASA}, summing an aperture radius of 15 pixels (3.75$^{\prime\prime}$ on sky), after subtracting the background per pixel. This was determined from the median value of a background regions located from 40--70 pixels either side of the centre of the spectral trace. Example spectra of the target and comparison stars are shown in Fig.~\ref{fig:spectra} for both grisms.

The cross-dispersion PSF was fitted for each pixel channel to obtain a time-series of the full-width at half-maximum (FWHM) and the $y$-position of the stars (averaged over wavelength). We also determined movement in the dispersion direction ($x$) by cross-correlating the spectra using the H$\beta$ line for the blue grism and the O$_2$ A band telluric feature for the red grism, after fitting a low-order polynomial to normalise the continua. The time-series spectra for each star were interpolated to the same wavelength scale using the $x$-shifts determined from the cross-correlation, and also from cross-correlation between the target and comparison star, again using the same spectral features.

\begin{figure}
\centering
\includegraphics[width=85mm]{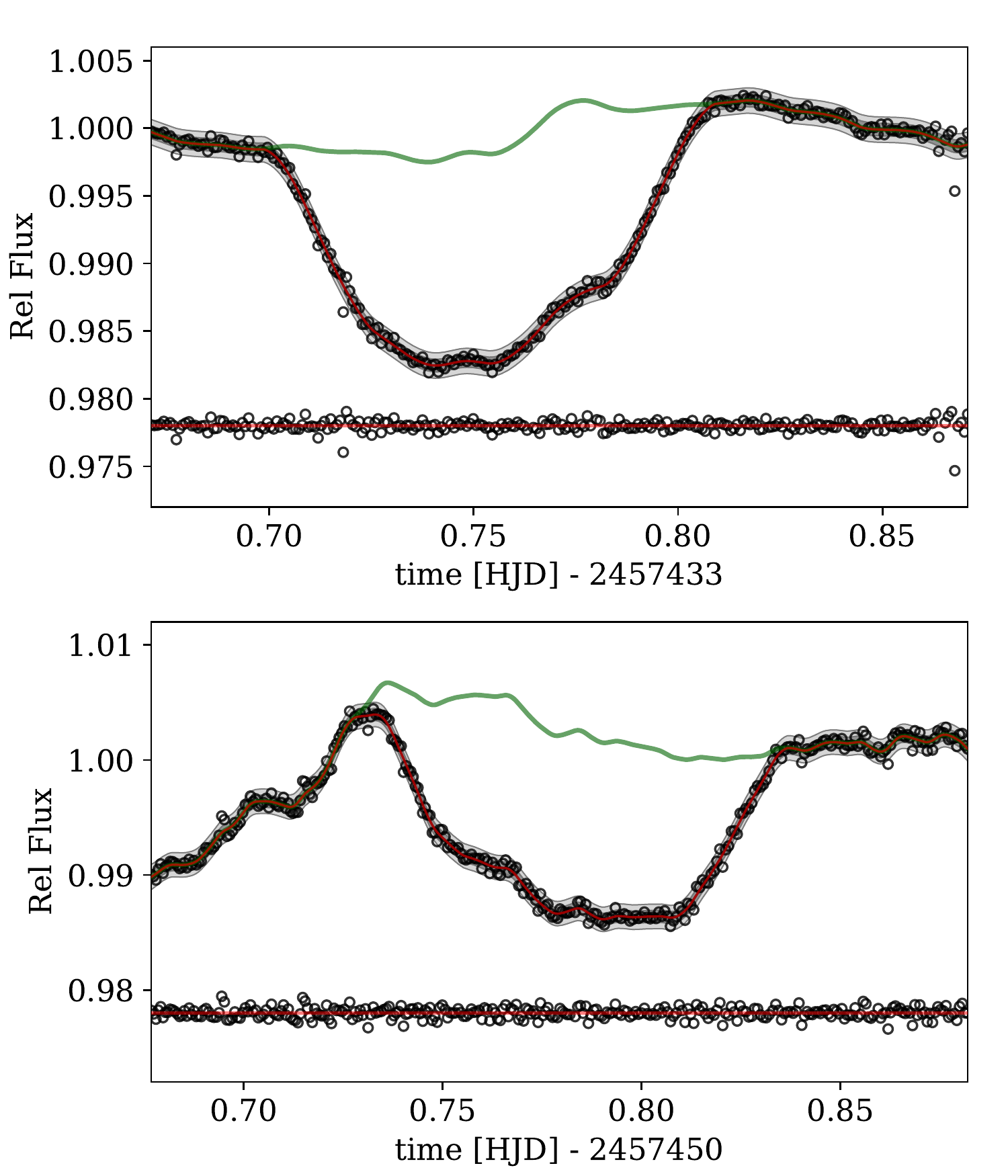}
\caption{`White' light curves of WASP-31b. The top and bottom panels show transits for the 600B and 600RI grisms, respectively. The grey shading indicates the mean plus/minus two standard deviations for the best fitting Gaussian process model. The red line shows the best fit model, and the green line shows the systematics models derived from the GP fits. The residuals are shown below each light curve.}
\label{fig:wlcs}
\end{figure}

We then proceeded to construct differential light curves from the time-series spectra. First, a `white' light curve was constructed for each transit by integrating the target and comparison stars' fluxes over a broad wavelength range, as indicated in Fig.~\ref{fig:spectra}, and dividing the time-series of the target star's flux by the comparison star's flux. We avoided using the edges of the spectra, and where the signal-to-noise was lower, although this has negligible influence on our results. We also extracted `spectral' light curves by integrating over narrower wavelength channels, again shown in Fig.~\ref{fig:spectra}, and discussed further in Sect.~\ref{sect:spec}. The white light curves and spectral light curves are shown in Figs.~\ref{fig:wlcs}, \ref{fig:speclcvs1} and \ref{fig:speclcvs2}. Finally, for the spectral response functions we extract limb darkening parameters using the {\sc PyLDTk} toolkit \citep{Parviainen2015}, which calculates limb darkening coefficients and uncertainties for a range of limb darkening laws using the spectral libraries of \citet{Husser2013}. We used the stellar parameters and uncertainties for WASP-31b derived in the discovery paper \citep{Anderson_2011}.

\section{Analysis}
\label{sect:analysis}
\subsection{White Light Curve Analysis}
\label{sect:wlc}

The white light curves and the spectral light curves exhibit large scale ($\sim$1 percent level) systematics for both transit datasets. Fortunately, the systematics are mainly `grey'; in other words to first order they do not vary with wavelength. We use this fact to correct the spectral light curves using a "common-mode correction" to high accuracy and therefore extract precise transmission spectra of the planet. A Gaussian process (GP) is used to model the systematics simultaneously with the deterministic transit model to untangle the signals. This is a similar procedure to that used in \citet{Gibson_2013a} and \citet{Gibson_2013b}. The primary difference is that the systematics are larger, and we therefore use prior information on the transit parameters in order to determine the systematics signal to higher accuracy. A similar process is used for our FORS2 observations described in \citet{Nikolov_2016}; however, the systematics were significantly smaller in this case, and a parametric model was used to construct the common-mode correction.

The exact cause of the systematics is unknown, but they are most likely due to remaining spatial inhomogeneities in the instrument/telescope throughput early in the optical path, as the underlying systematic variations tend to be largest when the telescope is rotating quickly, and they are strongly common-mode. It is worth noting however, that the target and comparison stars are located on different detectors. The most-likely cause is the LADC, which is unlikely to have a completely uniform throughput following removal of the anti-reflection coating. This is the only component of the instrument that rotates relative to the field. Other possibilities include variable vignetting of the field-of-view as the field rotates relative to the telescope structure, or scattered light, although we see no relationship between the size of the systematic variations and the phase/position of the moon. We also found no obvious correlations between the systematics and the auxiliary measurements (i.e. postition/width of the spectral trace, rotator angle/speed), and therefore fit the systematics using a time-correlated model.

We proceed by fitting each of the white transit light curves using our GP model. GPs were introduced to model stochastic signals in exoplanet time-series in \citet{Gibson_2012}, where the reader is referred for details. See \citet{Gibson_2013a,Gibson_2013b} for the application of GPs to similar datasets, and \citet{Gibson_2014} for a comparison to other commonly-used techniques for systematics analysis.\footnote{For a textbook level introduction see \citet{Rasmussen_Williams}, or \citet{Bishop} for a more general introduction in the context of Bayesian inference.}

Our GP is simply the multivariate Gaussian probability distribution defined as:
\[
p(\bmath f| \bmath t,\bphi,\btheta) = \mathcal{N} \left (T(\bmath t,\bphi) , \mathbf{\Sigma} (\bmath t,\btheta) \right).
\]
Here, $\bmath t$ and $\bmath f$ are vectors containing the time and flux measurements, respectively. $\mathcal{N}$ is the multivariate Gaussian, where the mean vector is described by a transit function $T(\bmath t,\bphi)$ with parameter vector $\bphi$. The covariance matrix $\mathbf{\Sigma} (\bmath t,\btheta)$ is defined by a covariance function or kernel, with parameters $\btheta$, and each element defined as
\[
\mathbf\Sigma_{nm} = k({t}_n,{t}_m | \btheta),
\]
for all times $n$ and $m$. As in \citet{Gibson_2013a,Gibson_2013b}, we use the Mat\'ern 3/2 kernel to model time-correlations in the data, given by:
\[
k({t}_n, {t}_m | \btheta) = \xi^2 \left( 1+{\sqrt{3}\,\eta\,\Delta t} \right) \exp \left( -{\sqrt{3}\,\eta\,\Delta t}\right) + \delta_{nm}\sigma^2,
\]
where $\xi$ is a hyperparameter that specifies the maximum covariance (and therefore the amplitude of the systematics), $\Delta t = |t_n-t_m|$ is the time difference, $\eta$ is the {\it inverse} length scale (defining the typical length scale of stochastic variations in the time-series), $\delta$ is the Kronecker delta, and $\sigma$ is the white noise component, assumed to be the same for all points in the time-series.

Our light curve model (i.e. the mean function) is calculated using the equations of \citet{Mandel_Agol_2002} using quadratic limb darkening (with parameters $c_1$ and $c_2$). We fitted for the central transit time ($T_c$), the system scale ($a/R_\star$), the planet-to-star radius ratio ($\rho = R_p /R_\star$), the impact parameter ($b$), and two parameters describing a linear function of time for the baseline ($f_{\rm oot}$, $T_{\rm grad}$). The period was held fixed, and Gaussian priors were placed on $a/R_\star$, $R_p /R_\star$, and $b$, using the values reported in \citet{Sing_2015}, or directly derived from them. We also placed Gaussian priors on the planet-to-star radius ratio, $\rho$, by determining the weighted mean reported over the wavelength ranges covered by the two white light curves. The limb darkening parameters were constrained using Gaussian priors with the best fit values and uncertainties from {\sc PyLDTk}. These are summarised in Tab.~\ref{tab:wlc_priors}. This naturally constrains the white light curve parameters to previously derived values, and enables a more accurate recovery of the common-mode systematics. The downside is that the overall level of the transmission spectrum is not derived from our data; however, this will not affect the conclusions of this study.

The kernel hyperparameters ($\xi$, $\eta$ and $\sigma$) were variable in the fitting process; however, we fitted for $\log\xi$ and $\log\eta$, which is equivalent to imposing priors of the form $p(x) = 1/x$, and is the natural choice of parameterisation for scale parameters. For example, fitting for the log length scale ($l = 1/\eta$) or log inverse length scale ($\eta$) are mathematically equivalent, i.e. this implies the same prior information, but this is not true when fitting for $l$ or $\eta$ directly. The length scale was also constrained to be no finer than the time-sampling, and no longer than the twice the full time-span using a truncated uniform prior in log space. This was to aid convergence of the fitting algorithm.

We fit for both of the white light curves simultaneously. To obtain the best fitting model we first optimised the posterior with respect to the transit and kernel hyperparameters using a differential evolution algorithm\footnote{as implemented in the {\sc SciPy} package, based on \citet{Storn_1997}}, and fine-tuned the best-fitting value with a Nelder-Mead simplex algorithm. We only use the best-fit model from the white light curves, so do not explore the posterior distribution to obtain uncertainty estimates.

We justified our use of the quadratic limb darkening law by comparing the light curves to those generated using the non-linear limb darkening law, using the limb darkening parameters generated from {\sc PyLDTk}. We found that for the white light curves and spectroscopic channels, the difference in transit depth was typically $<2\times10^{-5}$, and conclude that our choice of limb darkening law has little impact on our study.

Once we determined the best-fitting parameters, we extracted the mean of the Gaussian process conditioned on the observed data, and separated the systematic component from the transit model in order to generate a systematics-only model. This includes the linear baseline model. The white light curves and their best fit models are shown in Fig.~\ref{fig:wlcs}, along with their corresponding systematics models and residuals. We then proceed to model the spectroscopic light curves, using the systematics models derived here.

\begin{table}
\caption{Assumed values of the transit parameters for the white light curve fits. With the exception of the orbital period which was held fixed, Gaussian priors were placed on the parameters, using the mean and standard deviation given below.}
\label{tab:wlc_priors}
\begin{tabular}{rccc}
\hline
\noalign{\smallskip}
\smallskip
Grism & Parameter & Value \\
\hline
Both & $P$ & 3.405886 days (fixed) \\
~ & $a/R_\star$ &  $8.19\pm0.10$\\
~ & $b$ &  $0.761\pm0.018$ \\
600B & $\rho$ &  $0.12546\pm0.00026$ \\
~ & $c_1$ & $0.545\pm0.0017$ \\
~ & $c_2$ & $0.155\pm0.0025$ \\
600RI & $\rho$ &  $0.125054\pm0.00035$ \\
~ & $c_1$ & $0.413\pm0.0015$ \\
~ & $c_2$ & $0.159\pm0.0026$ \\
\hline
\end{tabular}
\end{table}

\subsection{Spectroscopic Light Curve Analysis}
\label{sect:spec}

\begin{figure*}
\centering
\includegraphics[width=150mm]{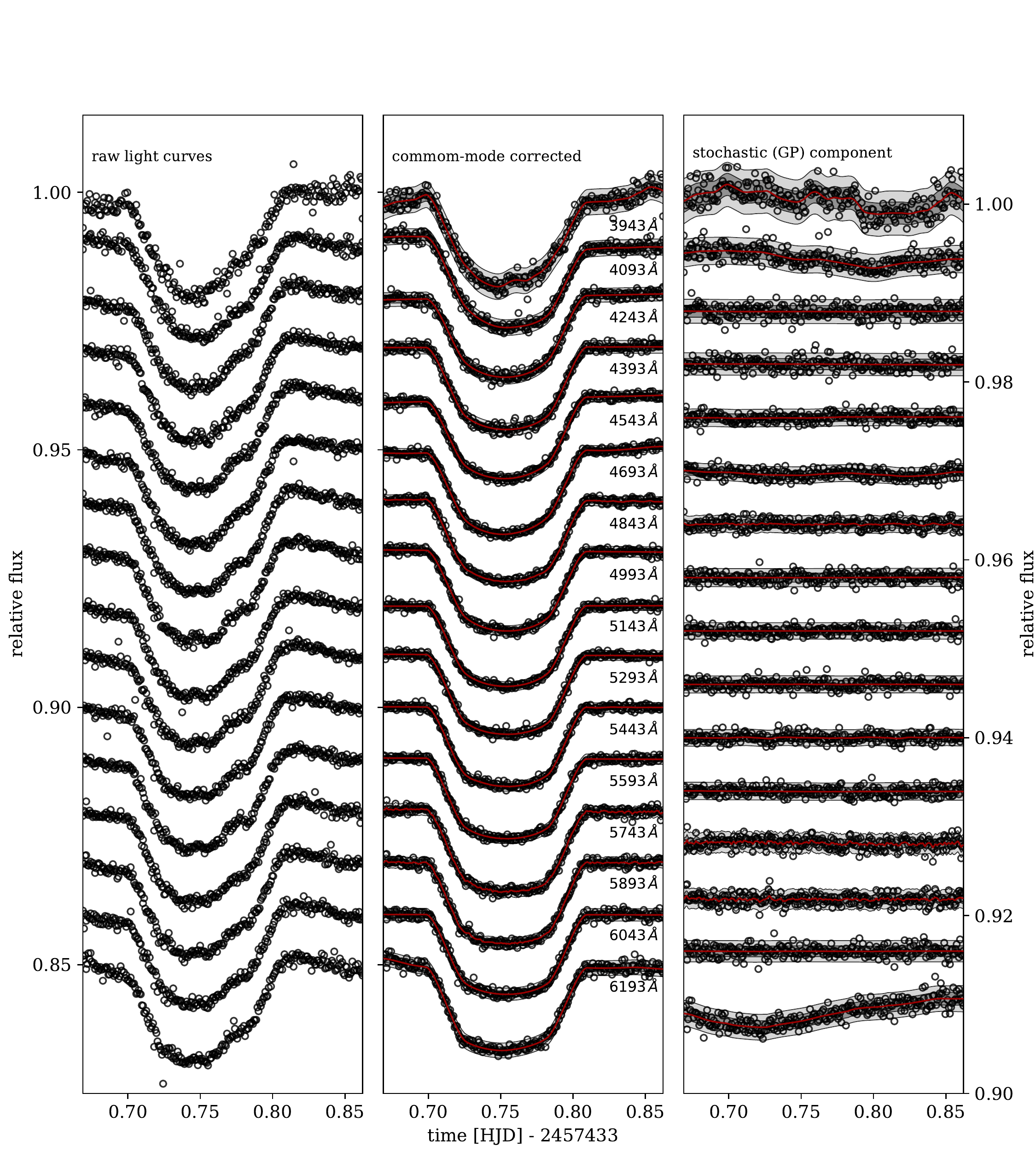}
\caption{Broad-band spectral light curves for the 600B grism. The left panel shows the raw light curves, after division by the comparison star's flux. The middle panel shows the common-mode corrected light curves, with best fit GP model, and the central wavelength of the spectral bin marked. The right panel shows the residuals from the best fit transit model, i.e. the stochastic component of the GP for the best-fitting hyperparameters.}
\label{fig:speclcvs1}
\end{figure*}

\begin{figure*}
\centering
\includegraphics[width=150mm]{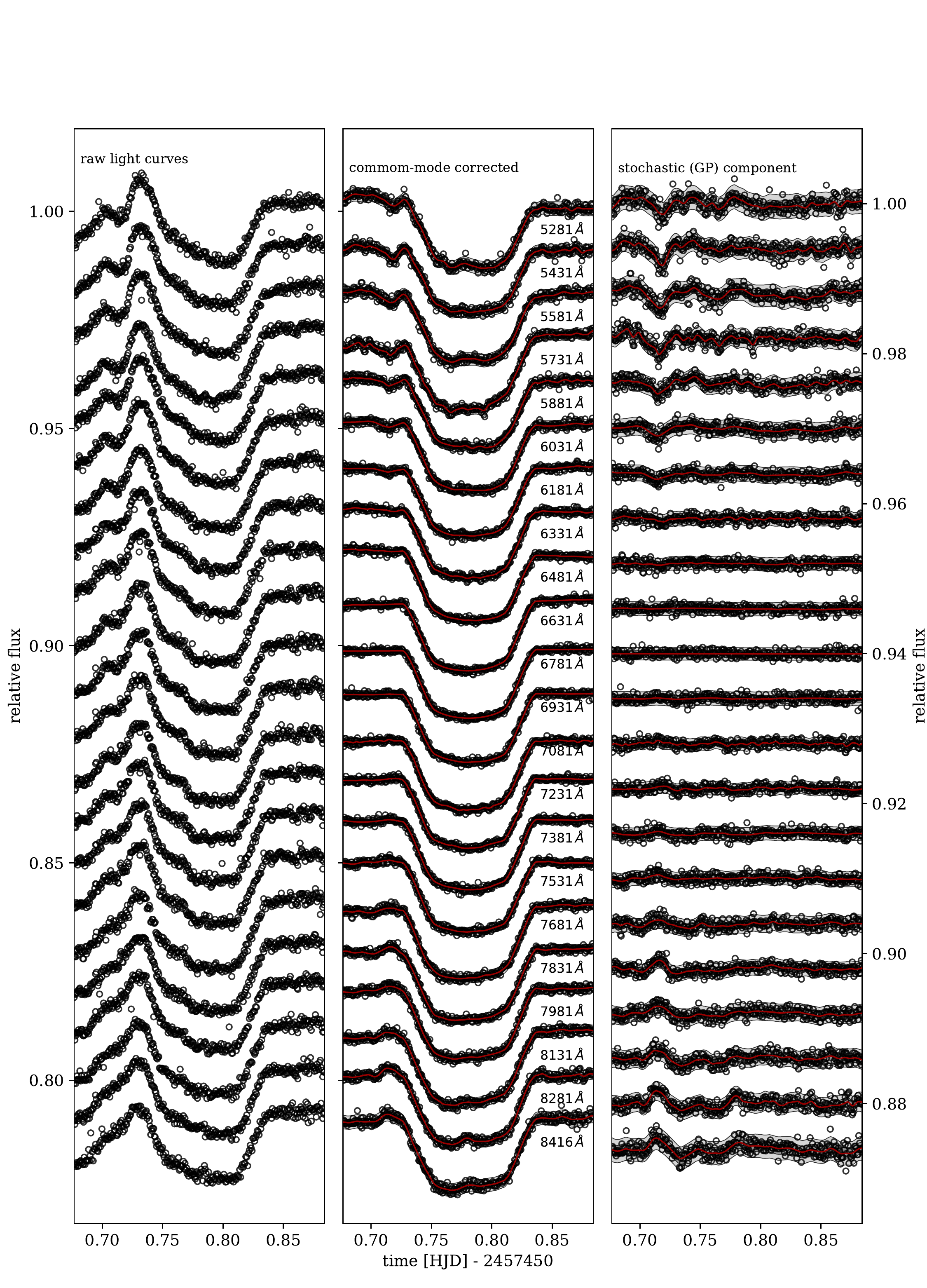}
\caption{Same as Fig.~\ref{fig:speclcvs1}, but for the broad-band spectral light curves for the 600RI grism.}
\label{fig:speclcvs2}
\end{figure*}

We first created low-resolution spectroscopic light curves, using uniform bins of width 150\,\AA, with the edges smoothed using a Tukey window of width 5\,\AA -- i.e. a cosine lobe. This was to mitigate the effect of sharp edges on the spectral channels. These correspond to the blue and red response curves shown in Fig.~\ref{fig:spectra}, which are multiplied by the target spectrum to show the spectral response for each channel. We extracted 16 of these low-resolution spectral channels for the 600B grism, and 22 for the 600RI grism. The resulting spectroscopic light curves are shown in Figs.~\ref{fig:speclcvs1} and \ref{fig:speclcvs2}. Clearly, they exhibit strong systematics, the main component of which is invariant in wavelength, and similar in shape to the white light curve systematics. We proceed by first removing these common-mode systematics by dividing each spectroscopic light curve by the systematics models derived from the respective white light curve fits. We also remove high-frequency systematics by subtracting the residuals from the white light curves and their best-fitting models. Such high-frequency systematics are often observed in such datasets, e.g. \cite{Gibson_2013a, Gibson_2013b}, and likely arise due to spatial variations in the atmospheric throughput (e.g. uneven cloud cover), or additional instrumental effects; however, it made little difference for these datasets. The results of this correction are also shown in Figs.~\ref{fig:speclcvs1} and \ref{fig:speclcvs2}.

The use of a common-mode correction enables us to recover higher-precision measurements of the relative planet-to-star radius ratio, and hence the transmission spectrum; however, it is important to note that it comes at the cost of (potentially) applying an offset to the overall depth of the transmission spectrum, as it assumes a single, best-fit systematics model. In other words our final posterior probability distributions are conditioned on a single common-mode correction, which has an associated uncertainty on the transit depth. The crucial (and reasonable) assumption is that the shape of the transmission spectrum does not change with different random draws of the systematics model. This is verified by finding consistent transmission spectra using different methods to find the common-mode correction, including excluding the priors from the white light curve fits. A full analysis would require marginalisation over transmission spectra generated using different common-mode corrections drawn from the posterior distribution of the white light curve fits, after correcting for offsets in the spectra; however, this is beyond the scope of this paper and is unlikely to impact our conclusions.

The uncertainty in the offset of the transmission spectrum is related to the uncertainty in the white light curve fits. In this case, we impose the overall depth of the transmission spectrum using our white light curve priors derived from HST/STIS data \citep{Sing_2015}, as discussed in Sect.~\ref{sect:wlc}. The absolute transit depth could still be derived in principle if the common-mode correction is not applied, but we would be marginalising the transit parameters over the same systematics signal, which would result in inflated error bars, and a highly correlated transmission spectrum.

\begin{figure*}
\centering
\includegraphics[width=160mm]{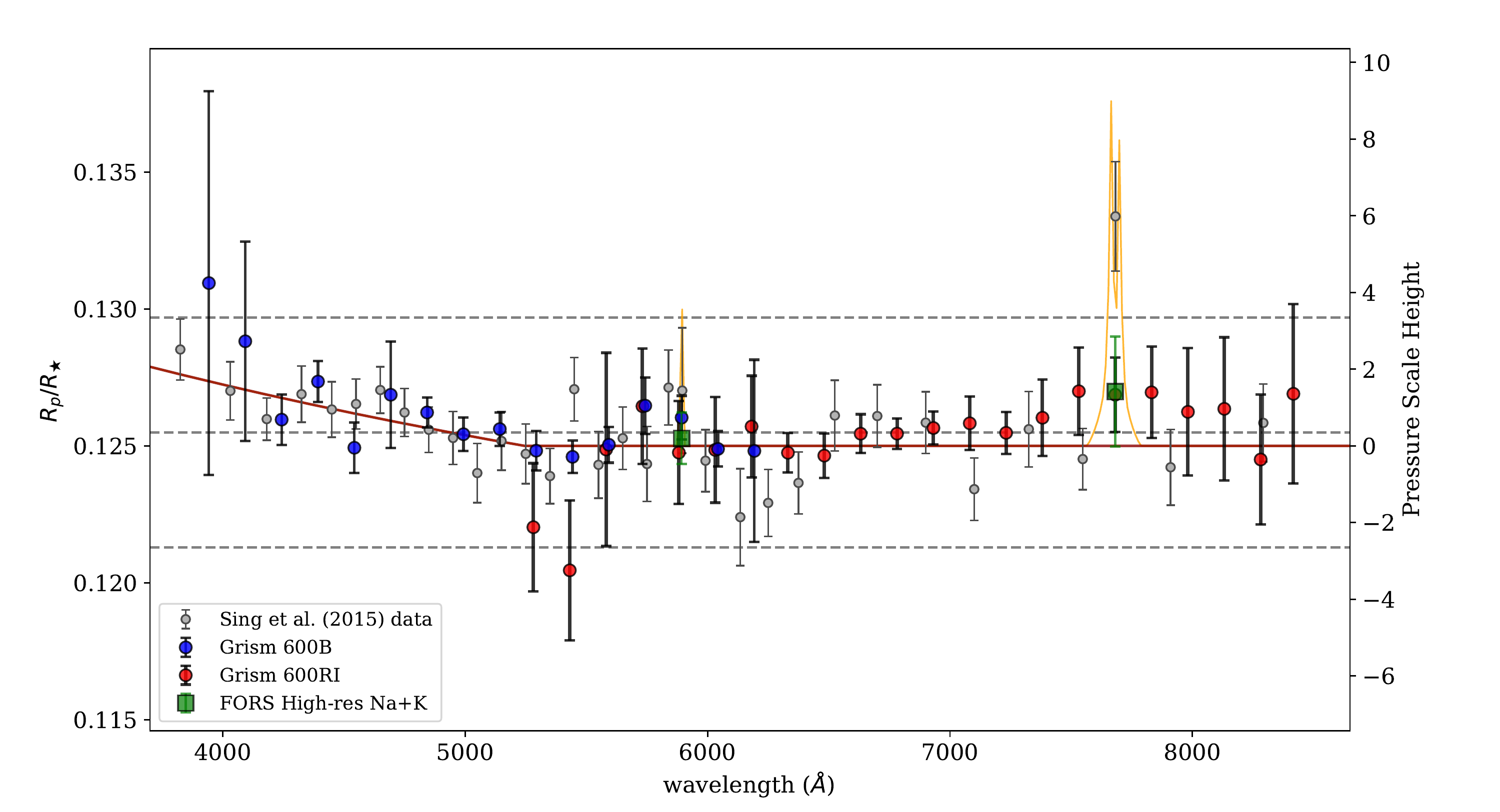}
\caption{FORS2 transmission spectrum of WASP-31b. The blue and red points are the results for the low-resolution light curves form the 600B and 600RI grisms, respectively. The grey points are the STIS results reported from \citet{Sing_2015}. The green squares are the high resolution channels extracted from the FORS2 data (binned for the two grisms for Na), as described in Sect.~\ref{sect:high_res}. The dashed lines correspond to the mean of the transmission spectrum, plus and minus 3 atmospheric scale heights. The solid lines correspond to the model reported in \citet{Sing_2015}, with (red) and without (orange) the Na and K features.
}
\label{fig:transpec}
\end{figure*}

We proceed to fit the common-mode corrected light curves with exactly the same model used for the white light curves; however, we fix the central transit time, the system scale, and the impact parameter at the value determined from the white light curve fits. To be conservative, rather than use the uncertainties from {\sc PyLDTk} as Gaussian priors, we arbitrarily increased these to have a standard deviation of 0.1. This did not have a major impact on our results. Again, we find the global maximum using a differential evolution algorithm. We then clip data points that are more than 4$\sigma$ from the best-fit GP model. This typically resulted in clipping only 1--2 points per transit, and the arbitrary nature of this procedure does not change our results. We then explore the posterior distribution using a Markov-Chain Monte-Carlo (MCMC) analysis, using the implementation described in \citet{Gibson_2014} and references therein. For each light curve we used 4 chains each of length 60,000, and discarded the first half of the chains. We checked for convergence using the Gelman and Rubin statistic \citep{GelmanRubin_1992}. The best-fit models are shown in Figs.~\ref{fig:speclcvs1} and \ref{fig:speclcvs2}, and the derived values for the planet-to-star radius ratio are shown in Tab.~\ref{tab:fors2_results} and plotted in Fig.~\ref{fig:transpec}.
Finally, we compared the fitted white noise values to the uncertainties calculated from the photon noise (excluding scintillation), finding these were typically 5-20\% larger than the theoretical calculations. However, this simple comparison neglects the contribution of the systematics to the error budget, which in this case depends on the covariance terms $\xi$ and $\eta$.

\begin{table}
\caption{Transmission spectra of WASP-31b from the FORS2 low-resolution spectroscopic light curves.}
\label{tab:fors2_results}
\begin{tabular}{cccc}
\hline
\noalign{\smallskip}
\smallskip
Wavelength & Radius ratio  & \multicolumn{2}{c}{Limb Darkening} \\
Centre [Range] (\AA) & ${R_p}/{R_\star}$ & c1 & c2 \\
\hline
\multicolumn{1}{l}{\it 600B} \\[1pt]
3943 [3868-4018] & $0.13094\pm0.00701$ & 0.763 & 0.025 \\[1pt]
4093 [4018-4168] & $0.12882\pm0.00364$ & 0.694 & 0.112 \\[1pt]
4243 [4168-4318] & $0.12596\pm0.00092$ & 0.740 & 0.053 \\[1pt]
4393 [4318-4468] & $0.12735\pm0.00075$ & 0.650 & 0.123 \\[1pt]
4543 [4468-4618] & $0.12493\pm0.00092$ & 0.644 & 0.127 \\[1pt]
4693 [4618-4768] & $0.12687\pm0.00195$ & 0.612 & 0.149 \\[1pt]
4843 [4768-4918] & $0.12622\pm0.00055$ & 0.558 & 0.170 \\[1pt]
4993 [4918-5068] & $0.12543\pm0.00061$ & 0.572 & 0.152 \\[1pt]
5143 [5068-5218] & $0.12562\pm0.00061$ & 0.555 & 0.149 \\[1pt]
5293 [5218-5368] & $0.12483\pm0.00072$ & 0.536 & 0.156 \\[1pt]
5443 [5368-5518] & $0.12460\pm0.00059$ & 0.520 & 0.155 \\[1pt]
5593 [5518-5668] & $0.12504\pm0.00066$ & 0.504 & 0.161 \\[1pt]
5743 [5668-5818] & $0.12647\pm0.00103$ & 0.490 & 0.164 \\[1pt]
5893 [5818-5968] & $0.12603\pm0.00080$ & 0.477 & 0.164 \\[1pt]
6043 [5968-6118] & $0.12489\pm0.00065$ & 0.465 & 0.163 \\[1pt]
6193 [6118-6268] & $0.12482\pm0.00333$ & 0.453 & 0.159 \\[1pt]
\hline
\multicolumn{1}{l}{\it 600RI} \\[1pt]
5282 [5206-5356] & $0.12203\pm0.00234$ & 0.539 & 0.154 \\[1pt]
5432 [5356-5506] & $0.12046\pm0.00256$ & 0.521 & 0.155 \\[1pt]
5582 [5506-5656] & $0.12487\pm0.00353$ & 0.507 & 0.160 \\[1pt]
5732 [5656-5806] & $0.12645\pm0.00210$ & 0.492 & 0.162 \\[1pt]
5882 [5806-5956] & $0.12476\pm0.00188$ & 0.478 & 0.163 \\[1pt]
6032 [5956-6106] & $0.12485\pm0.00193$ & 0.467 & 0.163 \\[1pt]
6182 [6106-6256] & $0.12571\pm0.00185$ & 0.454 & 0.157 \\[1pt]
6332 [6256-6406] & $0.12475\pm0.00072$ & 0.444 & 0.160 \\[1pt]
6482 [6406-6556] & $0.12465\pm0.00081$ & 0.421 & 0.165 \\[1pt]
6632 [6556-6706] & $0.12545\pm0.00071$ & 0.395 & 0.169 \\[1pt]
6782 [6706-6856] & $0.12545\pm0.00055$ & 0.412 & 0.157 \\[1pt]
6932 [6856-7006] & $0.12565\pm0.00061$ & 0.403 & 0.158 \\[1pt]
7082 [7006-7156] & $0.12583\pm0.00098$ & 0.394 & 0.155 \\[1pt]
7232 [7156-7306] & $0.12548\pm0.00077$ & 0.387 & 0.154 \\[1pt]
7382 [7306-7456] & $0.12603\pm0.00139$ & 0.378 & 0.152 \\[1pt]
7532 [7456-7606] & $0.12700\pm0.00160$ & 0.369 & 0.153 \\[1pt]
7682 [7606-7756] & $0.12687\pm0.00136$ & 0.361 & 0.152 \\[1pt]
7832 [7756-7906] & $0.12696\pm0.00167$ & 0.354 & 0.152 \\[1pt]
7982 [7906-8056] & $0.12624\pm0.00233$ & 0.348 & 0.151 \\[1pt]
8132 [8056-8206] & $0.12635\pm0.00261$ & 0.339 & 0.152 \\[1pt]
8282 [8206-8356] & $0.12450\pm0.00237$ & 0.328 & 0.150 \\[1pt]
8416 [8356-8476] & $0.12690\pm0.00327$ & 0.318 & 0.152 \\[1pt]
\hline
\end{tabular}
\end{table}

\subsection{Sodium and Potassium}
\label{sect:high_res}

\begin{figure*}
\centering
\includegraphics[width=150mm]{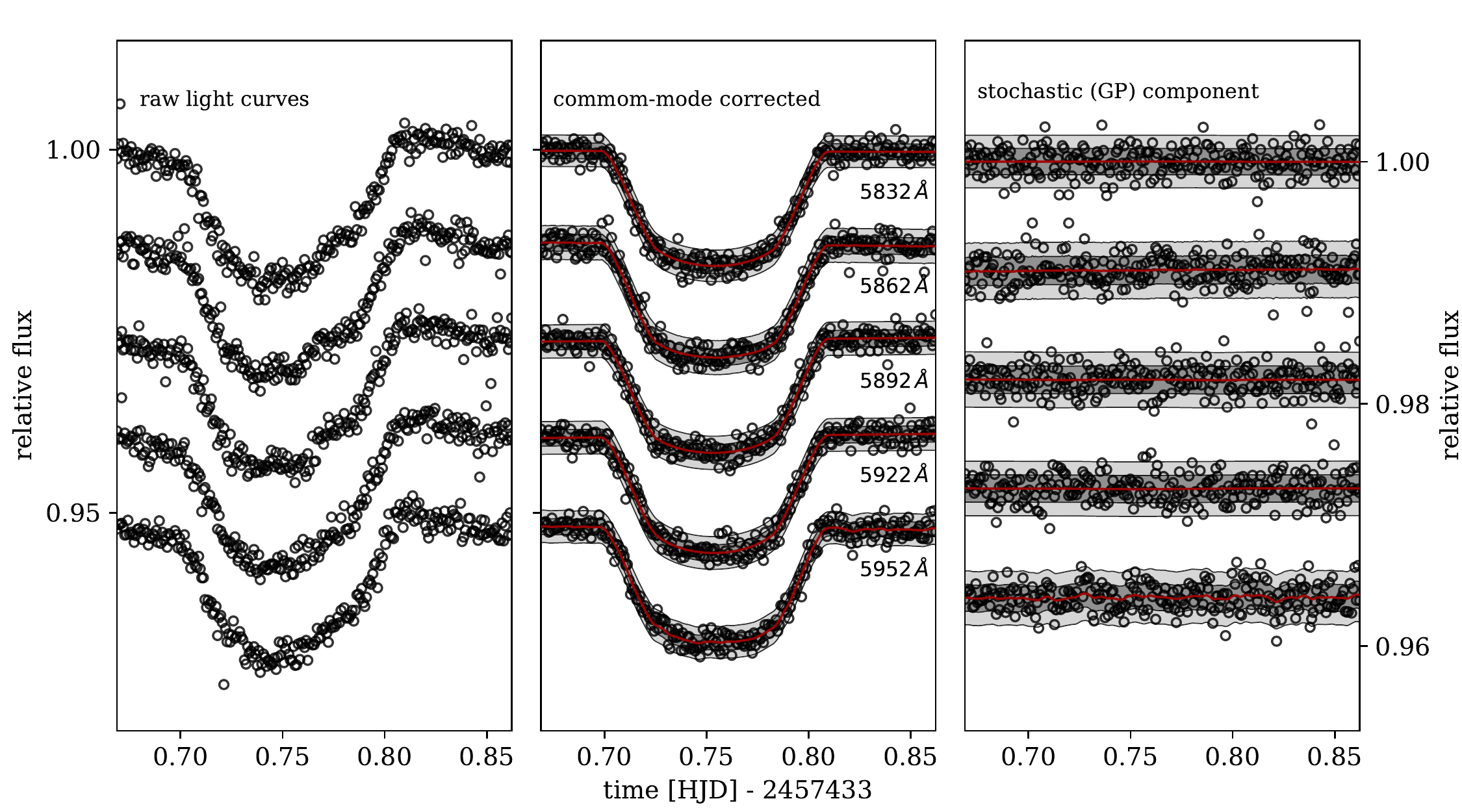}
\includegraphics[width=150mm]{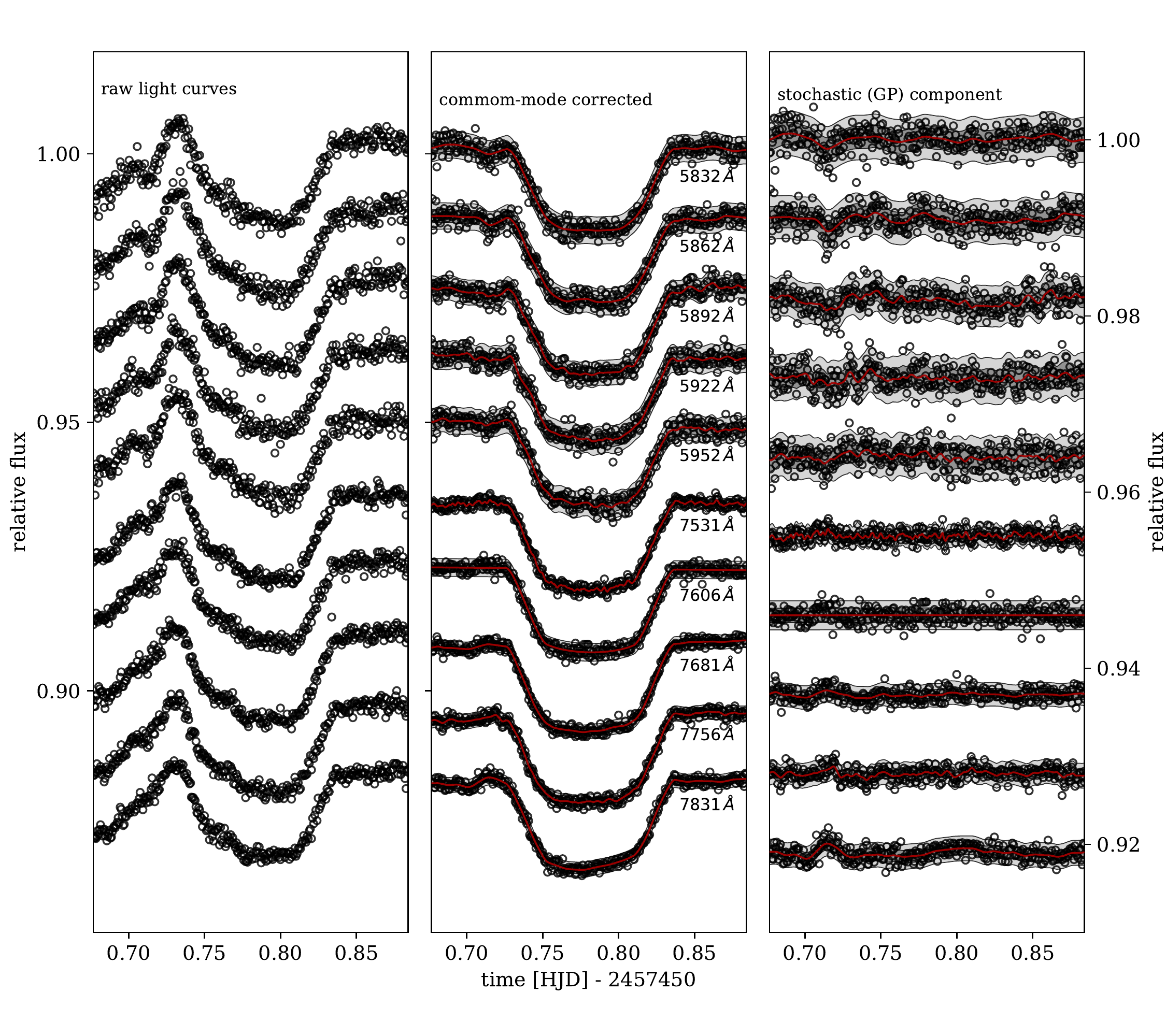}
\caption{Same as Figs.~\ref{fig:speclcvs1} and \ref{fig:speclcvs2}, showing the narrow-band spectral light curves of WASP-31b, centred around the Na and K features. The top and bottom plots show the 600B and 600RI grisms, respectively. The Na feature is covered using both grisms, whereas the K feature is only observed with the 600RI grism.}
\label{fig:speclcvs_HR}
\end{figure*}

In addition to the low-resolution light curves described in the previous section, we also search for signs of Na and K absorption by constructing high-resolution light curves. In particular this was motivated by trying to confirm the prominent K signature observed by \citet{Sing_2015}. For Na, we used 30\,\AA~bins centred at 5892.9\,\AA, and for K we used 75\,\AA~bins centred at 7681.5\,\AA. This corresponds to the same bins used in \citet{Sing_2015}. We also extracted light curves in the neighbouring continuum regions using the same bin widths, two at each side of each feature, in order to ensure the systematics were of similar amplitude. This resulted in sets of 5 higher-resolution light curves centred around the Na and K features. Both Na and K are covered by the 600RI grism, and only Na is covered by the 600B grism. These narrow spectral bins are shown in Fig.~\ref{fig:spectra}, and the corresponding light curves in Fig.~\ref{fig:speclcvs_HR}. The systematics follow the same functional form as for the wider channels, and we fitted the light curves using the same procedure as before, using the common-mode correction.

\begin{figure*}
\centering
\includegraphics[width=85mm]{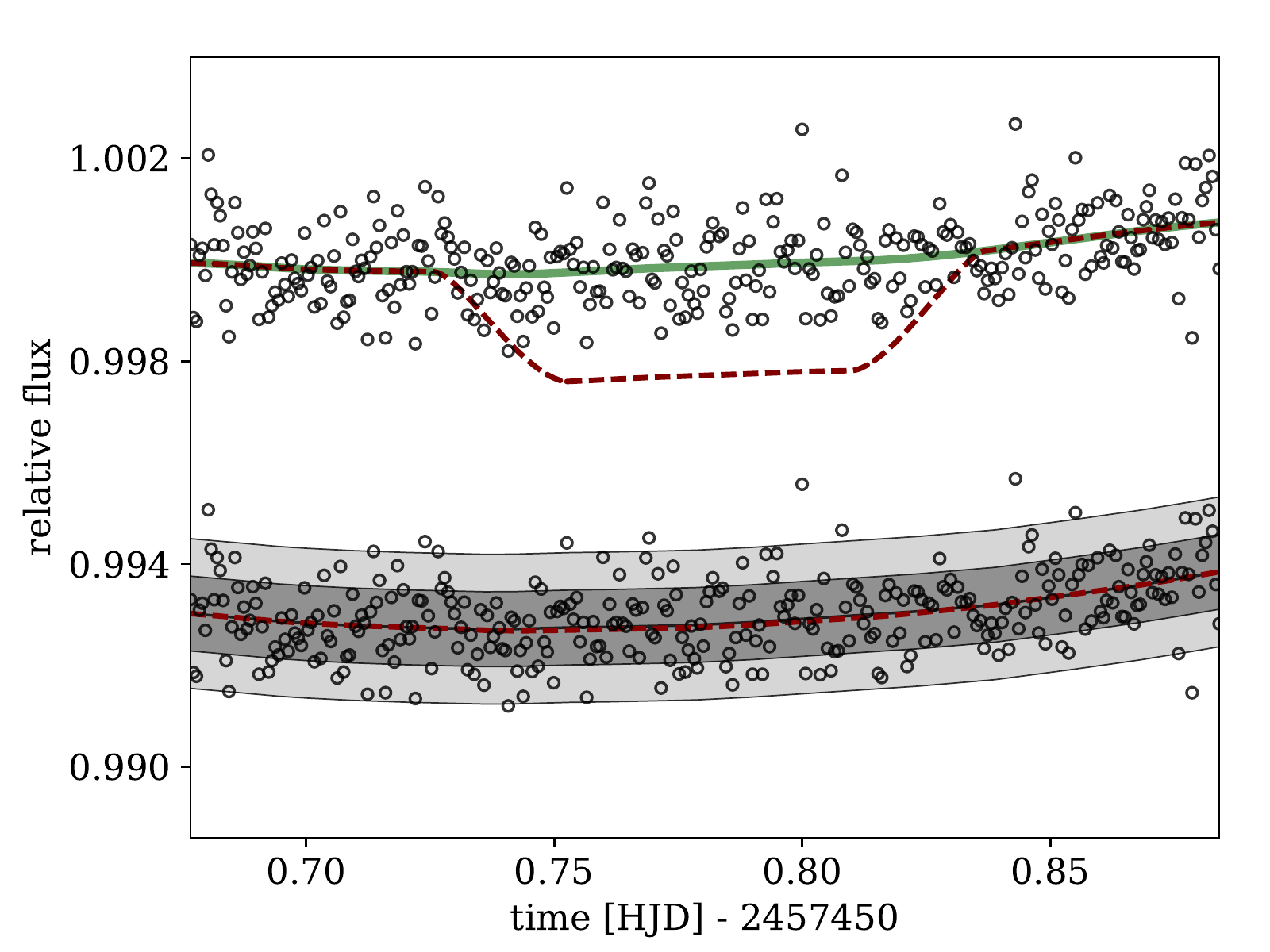}
\includegraphics[width=85mm]{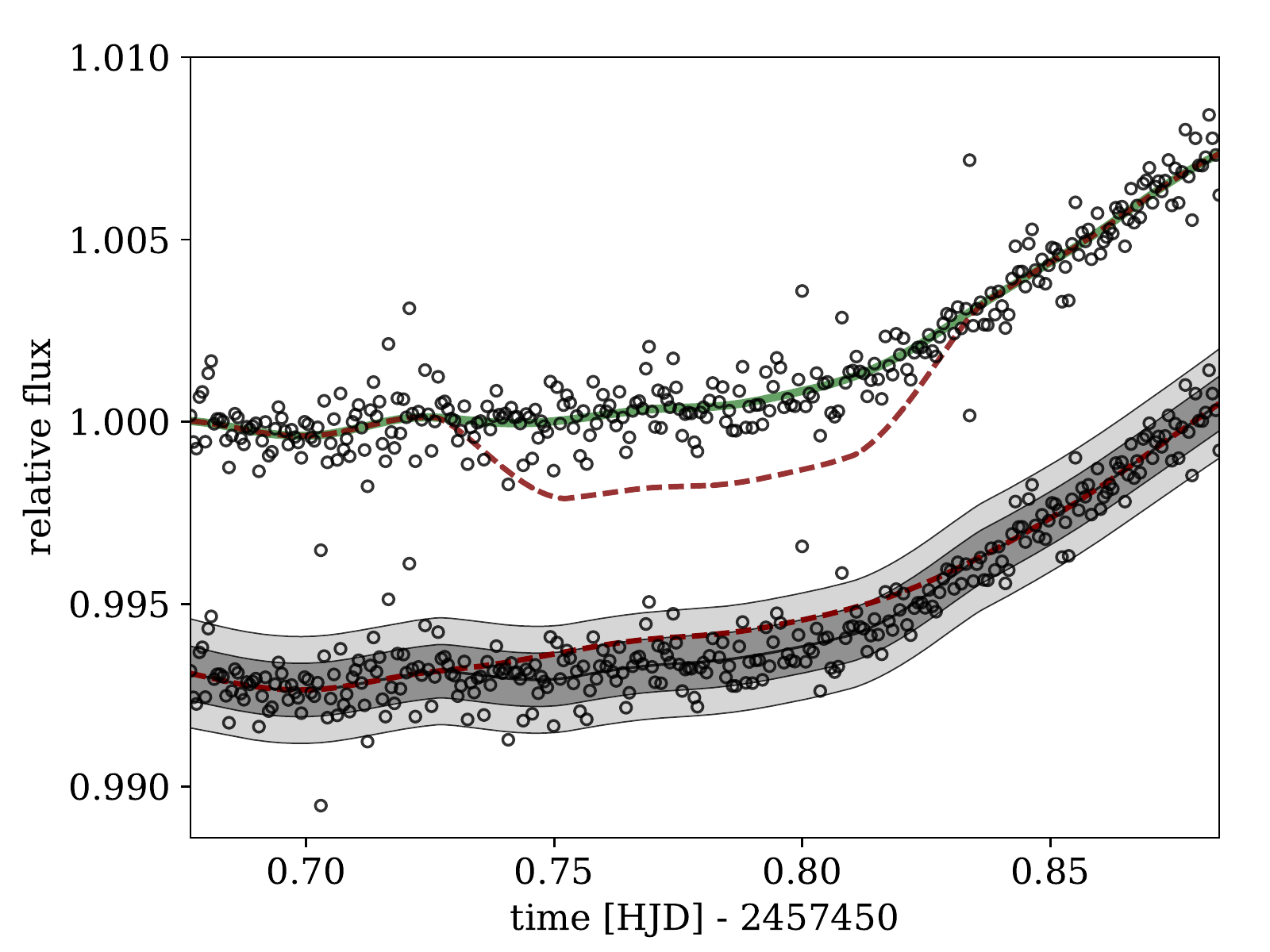}
\caption{Differential light curves centred on the narrow K channel. Left: light curve found by dividing the K channel differential light curve (ie. after correction for the comparison star) by the sum of the four differential continuum light curves. Right: light curve found by dividing the in-K {\it raw} light curves (ie. {\it without} correction for the comparison star) by the sum of the four raw continuum light curves. The upper plots show GP fits (green) to the light curves, with the expected signal from \citet{Sing_2015} overplotted. The bottom plots show GP fits including an eclipse model. No significant K feature is detected in either case.}
\label{fig:diff_speclcvs_HR}
\end{figure*}

The central Na (binned for both grisms) and K features are plotted in Fig.~\ref{fig:transpec}. No evidence for Na or K is found in our dataset. This is surprising as \citet{Sing_2015} reported significant K absorption, finding a planet-to-star radius ratio of 0.1334$\pm$0.0020, which should be easily detectable using our FORS2 data. We therefore decided to explore this further by constructing differential light curves around the K feature. First, we simply took the light curves for the central K band, and divided it through by the sum of the other 4 continuum light curves (i.e. those plotted in Fig.~\ref{fig:speclcvs_HR}). This is shown in the left of Fig.~\ref{fig:diff_speclcvs_HR}. We calculated the differential light curve depth from the \citet{Sing_2015} transmission spectrum, using the central K channel and the neighbouring 4 channels, finding $\Delta F = 0.00215\pm0.00053$. The dashed line in Fig.~\ref{fig:diff_speclcvs_HR} shows this signal over-plotted on the differential light curve, indicating that the signal should easily be detectable by eye in our light curves. The baseline (solid line) was determined using a GP fit to the light curve, using the same kernel as before, and a quadratic baseline in time as the mean function. The quadratic baseline was preferred by both the Bayesian Information Criterion \citep[BIC;][]{Schwarz_1978} and the Akaike Information Criterion \citep[AIC;][]{Akaike_1974}. To set an upper limit on the absorption, we also fitted for a differential transit as a mean function to the GP (we neglected the effects of differential limb darkening in this analysis). We found an eclipse depth of $\Delta F = -0.00006\pm0.00018$, indicating no detection of K, and we can place a 3$\sigma$ upper limit of $\Delta F\lesssim0.00048$ on the K absorption above the continuum. In addition, the AIC and BIC both favoured the model without an eclipse.

We also constructed a differential light curve ignoring the comparison star, i.e. by using only the {\it raw} flux from WASP-31 centred on the K feature, and the 4 neighbouring {\it raw} channels as reference to correct for the Earth's atmosphere. The light curve is shown on the right of Fig.~\ref{fig:diff_speclcvs_HR}, again alongside a GP fit and the projected differential eclipse assuming the K signal from \citet{Sing_2015}. There is clearly a larger variation in the baseline, due to the differential throughput of the Earth's atmosphere, particularly prominent as the K channel spans an O$_2$ telluric feature. Nonetheless, a visual inspection appears to rule out excess K absorption from WASP-31b. We performed a similar fit to determine an eclipse depth, finding $\Delta F = 0.00071\pm0.00042$, and setting a 3$\sigma$ upper limit on the K absorption of $\Delta F\lesssim0.002$. Again, the AIC and BIC favoured the model without an eclipse.

These results appear to rule out the large K absorption inferred from {\it HST}/STIS. However, it is important to emphasise that the K feature falls on a strong O$_2$ telluric feature that complicates detection of K from the ground. We note that other studies using FORS2 have reported problems extracting information around this telluric feature \citep{Sedaghati_2016}. In addition, our light curves suffer from significant instrumental systematics across the full spectral range, and we are assuming that the systematics are consistent for different narrow-band spectroscopic channels, and therefore cancel in the differential light curves. The interpretation of this result is discussed further in Sect.~\ref{sect:discussion}.

\begin{table}
\caption{Transmission spectra of WASP-31b centred on the Na and K features from the FORS2 high-resolution channels.}
\label{tab:fors2_results_HR}
\begin{tabular}{cccc}
\hline
\noalign{\smallskip}
\smallskip
Wavelength & Radius ratio  & \multicolumn{2}{c}{Limb Darkening} \\
Centre [Range] (\AA) & ${R_p}/{R_\star}$ & c1 & c2 \\
\hline
\multicolumn{1}{l}{\it 600B/Na} \\[1pt]
5833 [5818-5848] & $0.12629\pm0.00094$ & 0.482 & 0.167 \\[1pt]
5863 [5848-5878] & $0.12534\pm0.00095$ & 0.474 & 0.165 \\[1pt]
5893 [5878-5908] & $0.12522\pm0.00097$ & 0.481 & 0.160 \\[1pt]
5923 [5908-5938] & $0.12719\pm0.00104$ & 0.475 & 0.166 \\[1pt]
5953 [5938-5968] & $0.12651\pm0.00094$ & 0.473 & 0.164 \\[1pt]
\multicolumn{1}{l}{\it 600RI/Na} \\[1pt]
5833 [5818-5848] & $0.12466\pm0.00309$ & 0.482 & 0.166 \\[1pt]
5863 [5848-5878] & $0.12584\pm0.00279$ & 0.475 & 0.164 \\[1pt]
5893 [5878-5908] & $0.12612\pm0.00363$ & 0.481 & 0.159 \\[1pt]
5923 [5908-5938] & $0.12424\pm0.00192$ & 0.476 & 0.165 \\[1pt]
5953 [5938-5968] & $0.12360\pm0.00270$ & 0.473 & 0.162 \\[1pt]
\multicolumn{1}{l}{\it 600RI/K}  \\[1pt]
7532 [7494-7569] & $0.12727\pm0.00090$ & 0.369 & 0.153 \\[1pt]
7606 [7569-7644] & $0.12502\pm0.00070$ & 0.365 & 0.152 \\[1pt]
7682 [7644-7719] & $0.12698\pm0.00201$ & 0.360 & 0.152 \\[1pt]
7756 [7719-7794] & $0.12667\pm0.00101$ & 0.356 & 0.153 \\[1pt]
7832 [7794-7869] & $0.12755\pm0.00199$ & 0.355 & 0.151 \\[1pt]
\hline
\end{tabular}
\end{table}

\subsection{Re-analysis of HST/STIS data}
\label{sect:stis_analysis}

In order to resolve this contradiction in the K feature, and to produce a joint transmission spectrum, we re-analysed the {\it HST}/STIS data of WASP-31b that was first presented in \citet{Sing_2015}. These data were analysed using linear basis models to account for the systematics, using model selection via the BIC to choose the `correct' model. While this has proven to be a reliable technique for STIS analyses \citep[e.g.][]{Sing_2016,Nikolov_2016}, this method has also been shown to be the wrong statistical approach for determining transit parameters when faced with multiple systematics models, and consequently may underestimate the contribution of the choice of systematics models to the final transmission spectrum \citep{Gibson_2014}. We therefore reanalyse the data using a Gaussian process model to account for this additional source of uncertainty in the transmission spectrum, to test if this is a possible explanation for the strong K feature detected from {\it HST}/STIS, yet absent in our FORS2 data.

The STIS data consist of three transits of WASP-31b: two observed with the G430L grating, and one with the G750L grating. We refer the reader to \citet{Sing_2015} for further details, and comment only on the differences in the reduction applied here.

We started with the {\sc flt} images, which were processed using the latest version of the STIS pipeline ({\sc calstis v3.4}). This performed bias and flat-field corrections. The G750L data show significant fringing at long wavelengths, but we chose not to implement a fringing correction, as this is stable in time and therefore should not impact the light curves as long as the pointing is stable. Given the long exposure times (278 seconds), it was important to remove cosmic rays from the images. This was preformed by treating each pixel in the STIS array as a time-series. Each time-series was first divided through by the time-series averaged along each row (dispersion direction) of the array in order to remove the transit signal, and normalised. Outliers were identified by those more than 5 standard deviations from the median, and replaced by the value of a fitted interpolating function,\footnote{In this case we used a GP with a squared exponential kernel, but the exact nature of the interpolating function has little impact.} before being transformed back to the original counts level.

Spectral extraction was performed using a similar procedure to the FORS2 data, using the same aperture of 13 pixels as used by \citet{Sing_2015}, after background subtraction. We extracted the same diagnostic information as for the FORS2 data. However, we chose not to align the spectra to correct for movement in the dispersion direction, as the shift was sub-pixel. We checked that this didn't influence our final light curves. The white light curves were extracted as before, using a wide pass-band spanning the majority of the wavelength regions observed with STIS. The spectral light curves were extracted using the same spectral bins as \citet{Sing_2015} to enable easy comparison with these results, and also using the same bins as the FORS2 data to enable a combined analysis.

Rather than modelling the systematics purely as time-correlated noise, we also considered auxiliary inputs that describe the transient state of the observational setup. This is standard procedure for space based datasets \citep[e.g.][]{Brown_2001b,Gilliland_2003,Pont_2007,Sing_2011,Gibson_2011,Evans_2013,Nikolov_2014,Sing_2016}.
This is easy to perform using GPs, and we simply replace the time-dependent kernel used for the FORS2 data with the multidimensional kernel previously used in \citet{Gibson_2012,Gibson_2012b}:
\[
k(\bmath{x}_i, \bmath{x}_j) = \xi^2 \exp\left[-\sum_{k=1}^K\eta_k(x_{i,k} - x_{j,k})^2\right] + \delta_{ij} \sigma^2.
\]
$\xi$ and $\sigma$ again specify the maximum covariance and noise terms, respectively. $x_{i,k}$ is the $i$th element of the $k$th auxiliary input, and $\bmath\eta = \{\eta_1\dots\eta_K\}$ are inverse length scales, one for each of the $K$ input dimensions.

As inputs to the GP kernel, we used the orbital phase of {\it HST}, and the $x$- and $y$-positions of the spectral trace on the detector. These were included as they were the same inputs used by \citet{Sing_2015}. We also tested including the width and angle of the spectral trace, and found that they made negligible difference to the final spectra. We again chose to fit for $\log\xi$ and $\log\eta_k$. Fits were performed in the same way as before, first optimising the \mbox{(hyper-)parameters} using a global optimiser, and then using an MCMC to explore the joint posterior distribution. We again used four chains of length 60,000, and discarded the first half of the chains.

We first fitted the white light curve, allowing the kernel hyperparameters ($\log\xi$, $\log\eta_k$, $\sigma$) to vary along with $\rho$, $f_{\rm oot}$ and $T_{\rm grad}$. The values of $T_0$, $a/R_s$ and $b$ were fixed at those determined by \citet{Sing_2015}. The systematics component for each of the white light curves was isolated from the transit model, and used for the common-mode correction.

\begin{figure*}
\centering
\includegraphics[width=160mm]{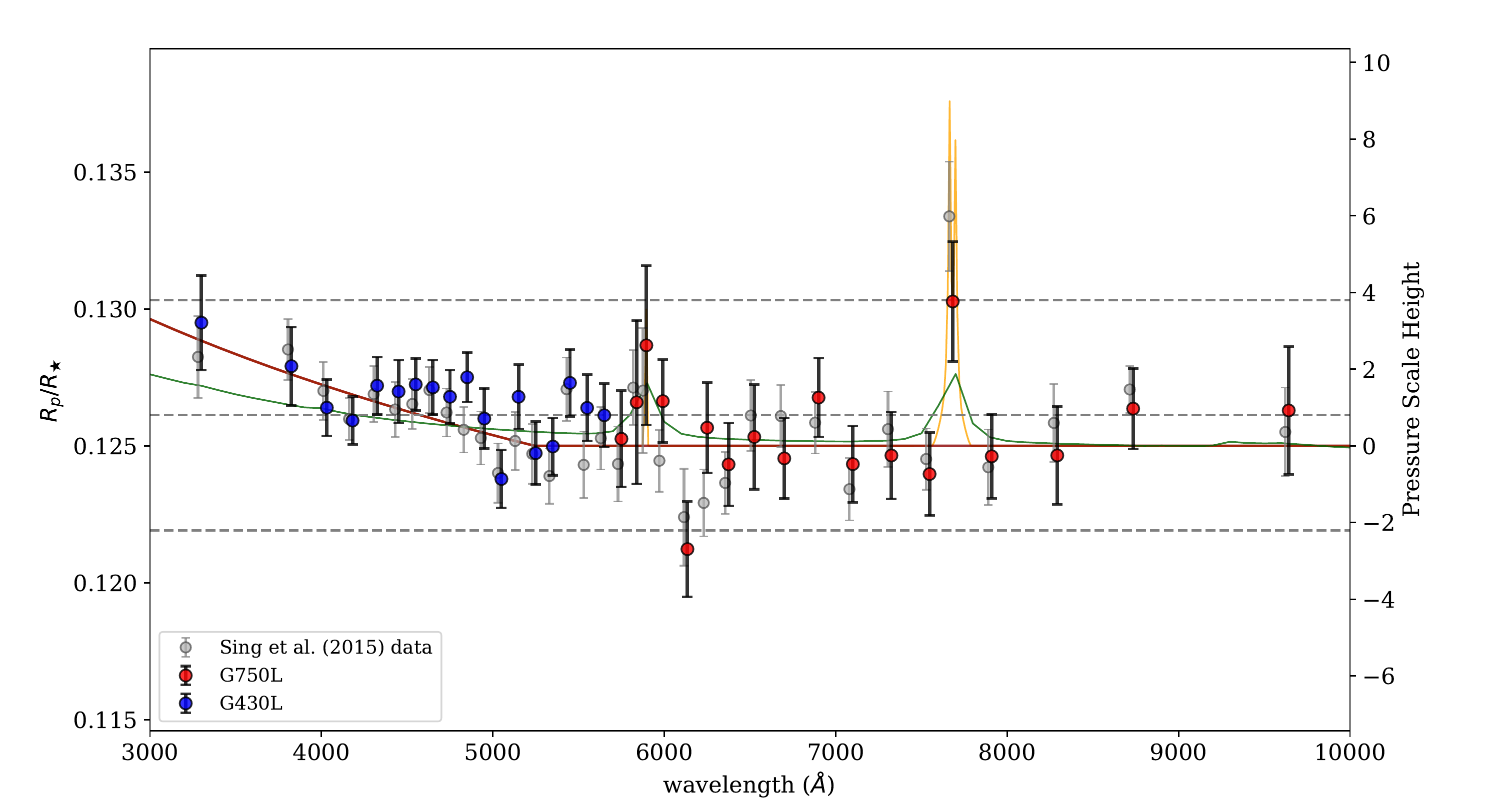}
\caption{Transmission spectrum of WASP-31b from the {\it HST}/STIS light curves. The blue and red points show the results for the G430L and G750L grisms, respectively. The grey points show the transmission spectrum from \citet{Sing_2015}, using the same data but a different systematics model. These are offset by 20\,\AA~for clarity. Our results are in excellent agreement, and the cloud-deck and Rayleigh scattering are reproduced. The K feature is also detected, although at reduced significance. The dashed lines correspond to the mean of the transmission spectrum, plus and minus 3 atmospheric scale heights. The solid orange and red lines correspond to the model reported in \citet{Sing_2015}, with and without the Na and K features, respectively, and the solid green line to the model reported in \citet{Barstow_2016}. Note that the latter is plotted at lower resolution, similar to that of the data.}
\label{fig:STIS_transpec}
\end{figure*}

The spectroscopic light curves were first divided through by the common-mode correction. We then fitted each of them using the same model, determining the planet-to-star radius ratio for each. The results are plotted in Fig.~\ref{fig:STIS_transpec}, using the same bins as the \citet{Sing_2015} analysis, and in Fig.~\ref{fig:joint_transpec}, using the same bins as the FORS2 data. Tab.~\ref{tab:stis_results} lists the results. The fitted white noise values for the spectroscopic light curves were on average $\sim10\%$ higher than the calculated photon noise for the two G430L transits, and $\sim15\%$ higher for the G750L transit, consistent with that reported in \citet{Sing_2015}. Again, we note that this neglects the contribution of systematics to the uncertainties.

\begin{table}
\caption{Transmission spectra of WASP-31b from the {\it HST}/STIS re-analysis.}
\label{tab:stis_results}
\begin{tabular}{cccc}
\hline
\noalign{\smallskip}
\smallskip
Wavelength & Radius ratio  & \multicolumn{2}{c}{Limb Darkening} \\
Centre [Range] (\AA) & ${R_p}/{R_\star}$ & c1 & c2 \\
\hline
\multicolumn{1}{l}{\it G430L} \\[1pt]
3300 [2900-3700] & $0.12950\pm0.00173$ & 0.802 & 0.036 \\[1pt]
3825 [3700-3950] & $0.12791\pm0.00143$ & 0.697 & 0.088 \\[1pt]
4032 [3950-4113] & $0.12640\pm0.00103$ & 0.703 & 0.099 \\[1pt]
4182 [4113-4250] & $0.12592\pm0.00087$ & 0.719 & 0.084 \\[1pt]
4325 [4250-4400] & $0.12720\pm0.00105$ & 0.694 & 0.080 \\[1pt]
4450 [4400-4500] & $0.12698\pm0.00114$ & 0.658 & 0.121 \\[1pt]
4550 [4500-4600] & $0.12725\pm0.00095$ & 0.648 & 0.123 \\[1pt]
4650 [4600-4700] & $0.12714\pm0.00100$ & 0.618 & 0.147 \\[1pt]
4750 [4700-4800] & $0.12679\pm0.00097$ & 0.604 & 0.149 \\[1pt]
4850 [4800-4900] & $0.12751\pm0.00091$ & 0.544 & 0.174 \\[1pt]
4950 [4900-5000] & $0.12600\pm0.00110$ & 0.579 & 0.149 \\[1pt]
5050 [5000-5100] & $0.12379\pm0.00105$ & 0.564 & 0.153 \\[1pt]
5150 [5100-5200] & $0.12679\pm0.00118$ & 0.556 & 0.145 \\[1pt]
5250 [5200-5300] & $0.12474\pm0.00115$ & 0.541 & 0.152 \\[1pt]
5350 [5300-5400] & $0.12498\pm0.00105$ & 0.532 & 0.154 \\[1pt]
5450 [5400-5500] & $0.12730\pm0.00122$ & 0.521 & 0.155 \\[1pt]
5550 [5500-5600] & $0.12639\pm0.00121$ & 0.508 & 0.160 \\[1pt]
5650 [5600-5700] & $0.12612\pm0.00116$ & 0.501 & 0.159 \\[1pt]
\hline
\multicolumn{1}{l}{\it G750L} \\[1pt]
5750 [5700-5800] & $0.12526\pm0.00175$ & 0.492 & 0.161 \\[1pt]
5839 [5800-5878] & $0.12659\pm0.00299$ & 0.481 & 0.163 \\[1pt]
5896 [5878-5913] & $0.12867\pm0.00291$ & 0.482 & 0.158 \\[1pt]
5992 [5913-6070] & $0.12663\pm0.00152$ & 0.471 & 0.161 \\[1pt]
6135 [6070-6200] & $0.12123\pm0.00174$ & 0.461 & 0.156 \\[1pt]
6250 [6200-6300] & $0.12566\pm0.00165$ & 0.449 & 0.158 \\[1pt]
6375 [6300-6450] & $0.12433\pm0.00152$ & 0.441 & 0.159 \\[1pt]
6525 [6450-6600] & $0.12533\pm0.00191$ & 0.392 & 0.173 \\[1pt]
6700 [6600-6800] & $0.12455\pm0.00147$ & 0.416 & 0.159 \\[1pt]
6900 [6800-7000] & $0.12676\pm0.00144$ & 0.405 & 0.158 \\[1pt]
7100 [7000-7200] & $0.12434\pm0.00140$ & 0.394 & 0.154 \\[1pt]
7325 [7200-7450] & $0.12465\pm0.00159$ & 0.381 & 0.153 \\[1pt]
7548 [7450-7645] & $0.12397\pm0.00151$ & 0.369 & 0.153 \\[1pt]
7682 [7645-7720] & $0.13027\pm0.00218$ & 0.360 & 0.152 \\[1pt]
7910 [7720-8100] & $0.12462\pm0.00154$ & 0.350 & 0.151 \\[1pt]
8292 [8100-8485] & $0.12465\pm0.00179$ & 0.327 & 0.151 \\[1pt]
8735 [8485-8985] & $0.12636\pm0.00147$ & 0.308 & 0.150 \\[1pt]
9642 [8985-10300] & $0.12630\pm0.00233$ & 0.292 & 0.149 \\[1pt]
\hline
\end{tabular}
\end{table}

\section{Discussion}
\label{sect:discussion}

WASP-31b is an inflated hot-Jupiter with mass and radius of $\approx$$0.48\,M_{\rm J}$ and $1.55\,R_{\rm J}$, respectively, a surface gravity 4.56\,m/s$^2$, and a zero-albedo equilibrium temperature of 1580\,K \citep{Anderson_2011}. This results in a scale height of 1220\,km, or 0.0014 $R_{\rm p}/R_\star$. \citet{Sing_2015} used nightly photometry over $\sim$3 years to monitor the activity of WASP-31, concluding that the effects of stellar variability are negligible in the interpretation of WASP-31b's transmission spectrum.

Using the {\it HST}/STIS data described previously, \citet{Sing_2015} detected a Rayleigh scattering signature at short wavelengths ($\lesssim550$\,nm), a cloud-deck (flat spectrum) at longer wavelengths, and a prominent K feature at 768\,nm. 
Following \citet{Sing_2015}, we interpret our transmission spectrum by fitting with a two component model including scattering at short wavelengths, and a flat spectrum at longer wavelengths. At short wavelengths we assume the scattering is dominated by a species with a power-law absorption cross-section with index $\alpha$: $\sigma =\sigma(\lambda/\lambda_0)^\alpha$. \citet{Lecavelier_2008a} showed that this results in a constant gradient with respect to $\ln\lambda$, given by
\[
\alpha T = \frac{\mu g}{k} \frac{dR_{\rm p}}{d\ln\lambda},
\]
where $T$ is the temperature of the atmosphere, $\mu$ is the mean molecular weight (assumed to be 2.3 times the proton mass), $g$ is the surface gravity, $k$ is the Boltzmann constant, and $R_{\rm p}$ is the planet's radius. At longer wavelengths we assume a flat spectrum, defined by the planetary radius given by the scattering law at a transition wavelength, $\lambda_T$.
\citet{Sing_2015} found $\alpha T = -9280\pm3240$\,K, and a transition wavelength $\lambda_T=510\pm30$\,nm.

\subsection{FORS2 Transmission Spectrum}

Our FORS2 transmission spectrum of WASP-31b is shown in Fig.~\ref{fig:transpec}. The broad-band spectrum is consistent with the results presented in \citet{Sing_2015}. We discuss the non-detection of K in the Sect.~\ref{sect:potassium}.
The uncertainties range from a factor of $\sim$2 lower than our STIS results at the centre of the FORS2 grisms (corresponding to $\approx1.4\times10^{-4}$), to a factor of $\sim$2 larger at the edges of the grisms ($\gtrsim1\times10^{-3}$). This reflects the importance of the common-mode correction to our FORS2 results, which is most accurate in the central wavelength bands, as well as the fact that the sensitivity falls off at the edges of the wavelength coverage, in particular at the bluest wavelengths covered by the 600B grism.

We use a Levenburg-Marquardt algorithm\footnote{as implemented in {\sc SciPy}} to fit the model transmission spectrum to the data, using a super-sampled model to integrate over each wavelength bin. We found $\alpha T = -7600\pm 5400$\,K, and a transition wavelength $\lambda_T=530\pm 40$\,nm. These are consistent with the results of \citet{Sing_2015}; however, owing to the lack of coverage at shorter wavelengths and the lower signal-to-noise at the edges of the wavelength coverage, the gradient at short wavelengths is not determined as precisely, and is only non-zero at a $\sim$1.4$\sigma$ significance.

We also fit the spectrum using a linear fit, and a flat spectrum. The model with the scattering signature and cloud-deck is preferred by the AIC ($\Delta$AIC = 0.1), whereas the flat model (cloud-deck only) is preferred by the BIC ($\Delta$ BIC = 1.2). Our FORS2 data therefore do not distinguish clearly between these two models. We also performed the same fits fixing the transition wavelength to 510\,nm, i.e. that determined by \citet{Sing_2015}. In this case both the AIC and BIC marginally prefer the two-component model ($\Delta$AIC = 1.1, $\Delta$BIC = 0.4), although again this is not a significant distinction.

The Rayleigh scattering slope is therefore only tentatively detected by the FORS2 data alone. This is unsurprising, as the sensitivity of FORS2 at the bluest wavelengths ($\lesssim$450\,nm) drops off sharply\footnote{This could be improved by using the blue-sensitive CCD, but this is only available in visitor mode.}, and the FORS2 data is of lower quality towards the extremes of each grism's wavelength coverage. Nonetheless, given the lack of broad Na and/or K wings predicted by cloud-free models \citep[e.g.][]{Fortney_2010}, we can unambiguously confirm the presence of scattering aerosols in the atmosphere, similarly to other ground-based observations of hot-Jupiters \citep[e.g.][]{Gibson_2013b,Nortmann_2016,Lendl_2016}.

\subsection{STIS Transmission Spectrum}

Our results from the re-analysis of the STIS data using GP models are shown in Fig.~\ref{fig:STIS_transpec}. They closely match those reported in \citet{Sing_2015}. The uncertainties are typically 10\% larger for the G430L grism, and 30\% larger for the G750L grism. This is explained by the extra flexibility allowed by the GP model, and the fact that it effectively marginalises over a large range of systematics models, rather than select a single one. Furthermore, our GP model assumes a {\it joint} systematics model of the inputs rather than independent signals summed together (as is the case for linear-basis models). Our model would be conceptually similar to adding cross-terms (e.g. $xy$, $x^2y$, etc) to the linear basis models, and marginalising over a large range of systematics models, although we note that our GP effectively marginalises over an infinite number of basis models \citep{Gibson_2012,Gibson_2014}.

Qualitatively, the Rayleigh scattering signature at wavelengths $\lesssim5500$\,\AA, and the cloud deck signature at longer wavelengths are reproduced. The models from \citet{Sing_2015} and \citet{Barstow_2016} are also plotted in Fig.~\ref{fig:STIS_transpec}, and our broad-band results remain consistent with these models. We did not reproduce the strong 4.3$\sigma$ K feature, but still detect an excess absorption in the narrow K-band. Using the neighbouring 4 points as reference, we determine the excess absorption in the narrow K feature to be $\Delta F=0.0015\pm0.0006$ ($\approx$2.5$\sigma$), indicating tentative evidence for K, but not a conclusive detection. This is likely due to the different systematics treatment presented here.

\subsection{Combined Broad-band Transmission Spectrum}

\begin{figure*}
\centering
\includegraphics[width=160mm]{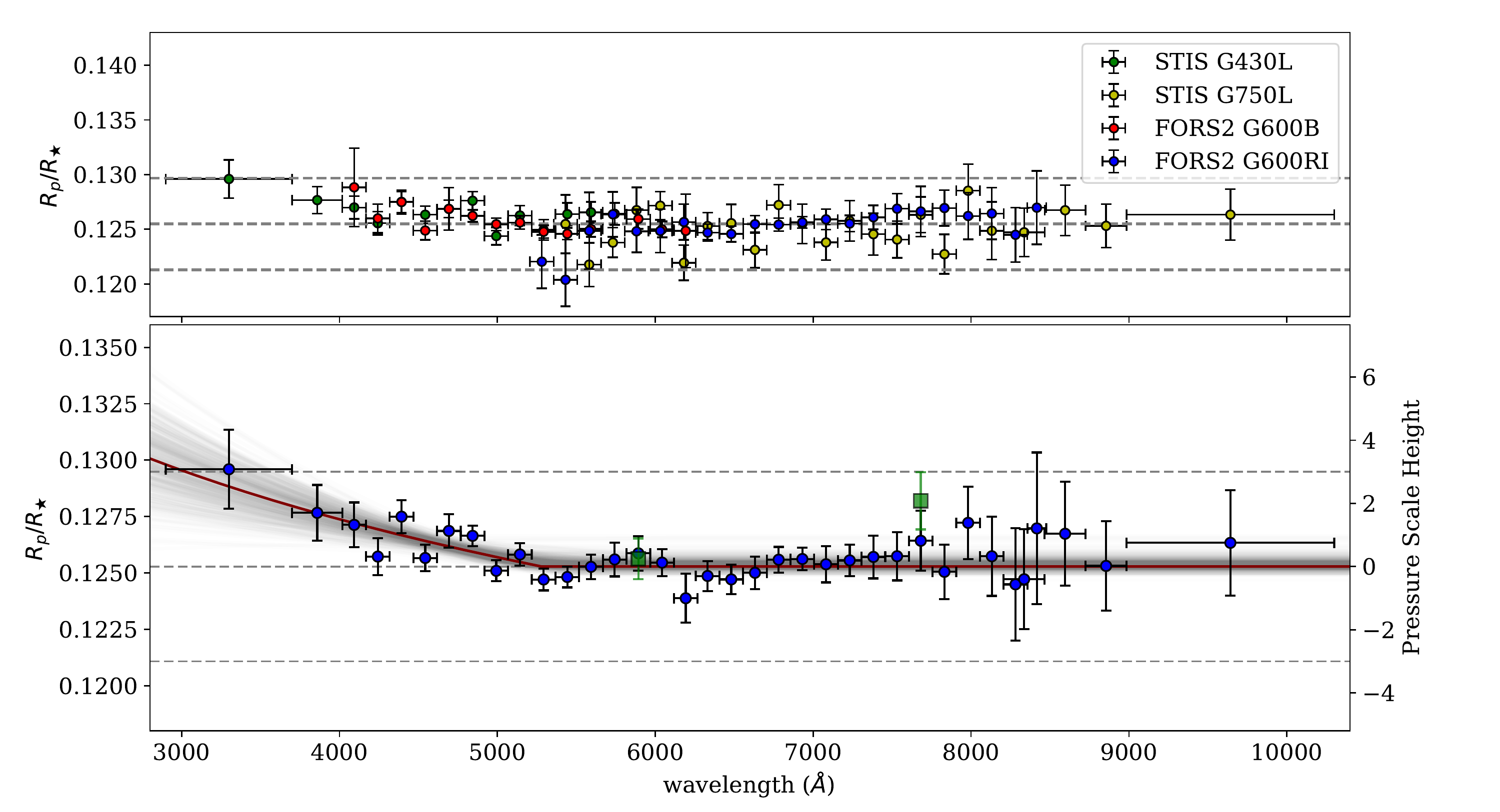}
\caption{Combined transmission spectrum using the FORS2 and STIS datasets. The upper panel shows the data for the individual grisms, and the lower panel is the combined dataset. The green squares are the high-resolution channels for Na and K (not shown in the upper plot). The horizontal error bars show the extent of the wavelength bins. These are not shown for the high-resolution Na and K bins, where the bin-widths are 30\,\AA~and 75\,\AA, respectively. The red line is the best-fitting two-component model (as described in Sect.~\ref{sect:discussion}). The grey lines show samples drawn from the model fit, illustrating the uncertainty in the model.}
\label{fig:joint_transpec}
\end{figure*}

The final transmission spectrum of WASP-31b is shown in Fig.~\ref{fig:joint_transpec}, after combining the FORS2 and STIS datasets using a weighted average where the bins overlap. The values are provided in Tab.~\ref{tab:joint_results}. We again fitted the two-component scattering plus cloud-deck model to the spectrum, finding $\alpha T = -8390\pm 2670$\,K, and a transition wavelength $\lambda_T=530\pm 30$\,nm. These values are consistent with \citet{Sing_2015}, although we measure a slightly shallower gradient at short wavelengths. The two-component model is strongly preferred over the flat model ($\Delta$AIC = 6.8, $\Delta$BIC = 6.2), indicating that our results confirm the interpretation of \citet{Sing_2015}, that of a scattering aerosol dominating the opacity at short wavelengths, with a cloud-deck dominating at longer wavelengths. The model provides a good fit to the data, giving $\chi^2=21.6$ with 33 degrees of freedom. Note that the narrow Na and K bins were not included in the fit.

Assuming the equilibrium temperature of 1580\,K, we find $\alpha=-5.31\pm1.69$, consistent with a Rayleigh scattering signature ($\alpha=-4$). Alternatively, assuming Rayleigh scattering, we find a temperature of $T = 2100\pm670$\,K, consistent with the equilibrium temperature, and also with temperatures where condensates might form in the atmosphere \cite[$\lesssim2000$\,K, e.g.][]{Fortney_2005}.

\subsection{The Potassium Feature}
\label{sect:potassium}

We also searched for Na and K absorption in the FORS2 data, using narrow bands centred on the respective doublets, as discussed in Sect.~\ref{sect:high_res}. We did not find evidence for Na or K detection in our FORS2 dataset, which is inconsistent with the 4.3$\sigma$ detection of K reported in \citet{Sing_2015}. The differential light curves discussed in Sect.~\ref{sect:high_res} and presented in Fig.~\ref{fig:diff_speclcvs_HR} provide further evidence that the K absorption is not present in our FORS2 data at the level measured by \citet{Sing_2015}. The interpretation of this is complicated by the K feature falling on a large O$_2$ telluric feature. In principal the comparison star should account for varying levels of telluric absorption in the atmosphere as the telescope tracks the target; however, it is often the case that systematics are larger both in telluric features, and where there are sharp boundaries in the target or comparison stars' spectra. It is nonetheless difficult to imagine a scenario where unaccounted-for systematics hide such a prominent K feature in the differential light curves, but do not appear elsewhere in the light curve -- in other words, the systematics would have to be closely correlated to the transit shape, and approximately match the amplitude of the K feature.

The O$_2$ feature also modifies the shape of the K bin, but again this is unlikely to dilute the signal enough to hide the K.
For broad K absorption from the planet this is apparent in the shape of the narrow spectral channels in Fig.~\ref{fig:spectra}. However, it is still possible that the K feature is concentrated in the narrow cores of each line of the doublet, that is not resolved in our spectral bins. To test this hypothesis we used the ESO {\sc SkyCalc} tool \citep{Noll_2014}, to compute the telluric absorption spectrum at high resolution for the extremes of airmass covered by our observations. It is possible that one of the lines of the doublet (but not both) overlaps with a deep telluric feature with absorption $\approx50\%$, depending on the exact barycentric correction. However, the resulting telluric features are extremely narrow, with FWHM $\approx0.04$\,\AA. Significant dilution of the K signal would require the K lines to exactly match up with telluric features, and be concentrated in similarly narrow cores. This scenario seems unlikely, as the STIS K feature is detected using a 75\,\AA~bin, and therefore such narrow features would be diluted by a large factor when integrating over a wide spectral bin, requiring a much larger planet-to-star radius ratio than reported at low-resolution. Even if this was the case for the one line in the doublet, this would dilute the signal by $\approx25\%$ which would still be detectable in our data. Furthermore, assuming the planet is tidally locked, the cores are likely to be wider than $\approx0.04$\,\AA~from rotational broadening alone.

We also re-analysed the {\it HST}/STIS data revealing an excess K absorption feature of $\Delta F=0.0015\pm0.0006$ ($\approx$\,$2.5\,\sigma$). This is of lower amplitude and with a marginally larger uncertainty than \citet{Sing_2015}, who reported $\Delta F=0.0022\pm0.0005$ ($\approx$$\,4.3\,\sigma$). Given the lower statistical significance, the K feature from the STIS data could be re-interpreted as a statistical fluctuation in the data, and therefore consistent with the lack of detection found with FORS2. The original detection could therefore be dismissed as an underestimate of the uncertainties for narrow band features.
Nonetheless, our re-analysis does still provide tentative evidence for excess K absorption, and it remains suspicious that the one place our ground-based data disagree with the original STIS analysis is close to a deep telluric signal.
Without a more detailed understanding of the telluric absorption (which is dependent on the unknown shape of the planet's K feature) we cannot completely dismiss it conspiring to hide the K, either from additional time-series systematics, dilution of the signal due to the narrow K cores overlapping with telluric lines, or indeed a combination of both.

Our combined transmission spectrum results in a K detection of $\approx$$\,2.2\,\sigma$, therefore does not confirm the interpretation of excess K absorption found in \citet{Sing_2015}, although neither can it rule it out.
The interpretation of our results largely depends on the choice between the Linear Basis Model or GP approach, and on the effect of the O$_2$ telluric feature on our ability to detect K using our FORS2 data, which we cannot definitively quantify. We therefore cannot completely rule out K absorption at the level reported by \citet{Sing_2015}, although our FORS2 results and STIS re-analysis do raise questions about this interpretation.

\begin{table}
\caption{Combined transmission spectra of WASP-31b from the FORS2 and {\it HST}/STIS observations.}
\label{tab:joint_results}
\begin{tabular}{cccc}
\hline
\noalign{\smallskip}
\smallskip
Wavelength & Radius ratio  & \multicolumn{2}{c}{Limb Darkening} \\
Centre [Range] (\AA) & ${R_p}/{R_\star}$ & c1 & c2 \\
\hline
3300 [2900-3700] & $0.12960\pm0.00176$ & 0.798 & 0.041 \\[1pt]
3859 [3700-4018] & $0.12766\pm0.00124$ & 0.703 & 0.086 \\[1pt]
4093 [4018-4168] & $0.12713\pm0.00100$ & 0.693 & 0.113 \\[1pt]
4243 [4168-4318] & $0.12572\pm0.00083$ & 0.740 & 0.053 \\[1pt]
4393 [4318-4468] & $0.12749\pm0.00073$ & 0.649 & 0.124 \\[1pt]
4543 [4468-4618] & $0.12566\pm0.00058$ & 0.644 & 0.128 \\[1pt]
4693 [4618-4768] & $0.12686\pm0.00074$ & 0.612 & 0.149 \\[1pt]
4843 [4768-4918] & $0.12664\pm0.00046$ & 0.558 & 0.170 \\[1pt]
4993 [4918-5068] & $0.12509\pm0.00047$ & 0.572 & 0.151 \\[1pt]
5143 [5068-5218] & $0.12582\pm0.00050$ & 0.555 & 0.149 \\[1pt]
5293 [5218-5368] & $0.12471\pm0.00048$ & 0.621 & 0.120 \\[1pt]
5443 [5368-5518] & $0.12482\pm0.00046$ & 0.607 & 0.134 \\[1pt]
5593 [5518-5668] & $0.12527\pm0.00054$ & 0.623 & 0.107 \\[1pt]
5743 [5668-5818] & $0.12560\pm0.00075$ & 0.570 & 0.143 \\[1pt]
5893 [5818-5968] & $0.12587\pm0.00077$ & 0.560 & 0.146 \\[1pt]
6043 [5968-6118] & $0.12546\pm0.00061$ & 0.539 & 0.156 \\[1pt]
6193 [6118-6268] & $0.12388\pm0.00109$ & 0.506 & 0.164 \\[1pt]
6332 [6256-6406] & $0.12486\pm0.00066$ & 0.443 & 0.160 \\[1pt]
6482 [6406-6556] & $0.12471\pm0.00066$ & 0.420 & 0.166 \\[1pt]
6632 [6556-6706] & $0.12501\pm0.00072$ & 0.394 & 0.169 \\[1pt]
6782 [6706-6856] & $0.12559\pm0.00057$ & 0.411 & 0.158 \\[1pt]
6932 [6856-7006] & $0.12562\pm0.00050$ & 0.402 & 0.159 \\[1pt]
7082 [7006-7156] & $0.12538\pm0.00080$ & 0.393 & 0.155 \\[1pt]
7232 [7156-7306] & $0.12555\pm0.00069$ & 0.386 & 0.155 \\[1pt]
7382 [7306-7456] & $0.12571\pm0.00095$ & 0.377 & 0.153 \\[1pt]
7532 [7456-7606] & $0.12574\pm0.00107$ & 0.369 & 0.154 \\[1pt]
7682 [7606-7756] & $0.12643\pm0.00133$ & 0.360 & 0.152 \\[1pt]
7832 [7756-7906] & $0.12505\pm0.00121$ & 0.354 & 0.152 \\[1pt]
7982 [7906-8056] & $0.12722\pm0.00160$ & 0.347 & 0.151 \\[1pt]
8132 [8056-8206] & $0.12574\pm0.00176$ & 0.338 & 0.152 \\[1pt]
8282 [8206-8356] & $0.12449\pm0.00249$ & 0.328 & 0.150 \\[1pt]
8416 [8356-8476] & $0.12697\pm0.00337$ & 0.318 & 0.152 \\[1pt]
8336 [8206-8466] & $0.12472\pm0.00221$ & 0.323 & 0.152 \\[1pt]
8596 [8466-8726] & $0.12674\pm0.00231$ & 0.310 & 0.149 \\[1pt]
8855 [8726-8985] & $0.12531\pm0.00199$ & 0.306 & 0.153 \\[1pt]
9642 [8985-10300] & $0.12633\pm0.00233$ & 0.291 & 0.150 \\[1pt]
\hline
\multicolumn{3}{l}{\it high-resolution channels$^\star$} \\[1pt]
5893 [5878-5908] & $0.12562\pm0.00090$ & 0.481 & 0.159 \\[1pt]
7682 [7644-7719] & $0.12820\pm0.00127$ & 0.359 & 0.152 \\[1pt]
\hline
\end{tabular}
\end{table}

\section{Conclusion}
\label{sect:conclusion}

We have presented a ground-based transmission spectrum of the hot-Jupiter WASP-31b using two transits to cover the wavelength range $\approx400-840\,$nm, using the recently upgraded FORS2 instrument on the VLT. This is the second paper in a series to re-observe targets already studied in detail with {\it HST}, both to confirm the signals detected with {\it HST}, and to verify the use of FORS2 for transmission spectroscopy. The results for our first target, WASP-39b, were presented in \citet{Nikolov_2016}. We also presented a re-analysis of the {\it HST}/STIS transmission spectrum of WASP-31b presented in \citet{Sing_2015}. 

Our FORS2 observations suffer from significant systematic effects, showing $\sim1$ percent level variations in both the broad-band and spectroscopic light curves. The exact cause(s) of the systematics is unknown, but the most likely culprit is the LADC, or some other spatial variation in the throughput early in the optical path. Luckily, the observed systematics are predominantly common-mode, i.e. they are invariant with wavelength to first order. We use this fact to extract a systematics model from the broad-band (`white') light curves, and use this to correct the spectroscopic light curves prior to extracting the transmission spectrum.

The FORS2 observations are consistent with the {\it HST}/STIS results and confirm the presence of a cloud-deck in the atmosphere of WASP-31b, owing to the flat spectrum across most of the optical range, and the lack of broad Na and/or K pressure broadened wings. We do not unambiguously detect the presence of Rayleigh scattering from the FORS2 observations alone. We also cannot reproduce the large K absorption reported in \citet{Sing_2015}, either using high-resolution channels in the transmission spectrum, or from differential light curves. As the K feature falls within a strong telluric O$_2$ band, the FORS2 data do not conclusively rule out a strong K core in WASP-31b.

We also re-analysed the {\it HST}/STIS data using a Gaussian process model, finding consistent results with \citet{Sing_2015}, only with slightly larger uncertainties. Our combined analysis confirms the overall picture from \citet{Sing_2015}, that of a Rayleigh scattering haze dominating the opacity at short wavelengths, and a cloud-deck dominating at longer wavelengths.
Our GP model detects excess absorption around the K feature, although with a larger uncertainty and slightly lower amplitude, resulting in a reduced detection significance of $\approx$\,2.5\,$\sigma$, and $\approx$\,2.2\,$\sigma$ when combined with the FORS2 data.

The combined FORS2 and STIS transmission spectrum therefore only shows tentative evidence for K absorption, and the low-significance detection from the STIS data could be interpreted as a statistical fluctuation. We therefore cannot confirm the sub-solar Na/K abundance inferred by \citet{Sing_2015}. However, the interpretation of our narrow-band results depends on the choice of systematics models applied to the STIS data (Linear-basis models vs Gaussian processes), and on the impact of the O$_2$ telluric absorption on the K feature, which is difficult to definitively quantify. Our results therefore raise some doubts on the interpretation of narrow-band features in low-resolution spectra of hot-Jupiters.  Nonetheless, the overall agreement of the broad-band FORS2 and STIS spectra, combined with our previous results from \citet{Nikolov_2016}, is a powerful demonstration of the stability of both instruments, and in our current interpretation of scattering signatures in hot-Jupiters \citep[e.g.][]{Sing_2016}. Resolving the discrepancy of the K feature will require further observations, preferably at higher-resolution.

\section*{Acknowledgements}

This work is based on observations collected at the European Organisation for Astronomical Research in the Southern Hemisphere under ESO programme 096.C-0765. N. P. G. gratefully acknowledges support from the Royal Society in the form of a University Research Fellowship. N. N, D. K. S, and T. M. E. acknowledge funding from the European Research Council under the European Unions Seventh Framework Programme (FP7/2007-2013) / ERC grant agreement no. 336792. J. K. B. is supported by a Royal Astronomical Society Research Fellowship. P.A.W. acknowledges the support of the French Agence Nationale de la Recherche (ANR), under program ANR-12-BS05-0012 `Exo-Atmos'. We are grateful to the developers of the {\sc NumPy, SciPy, Matplotlib, iPython} and {\sc Astropy} packages, which were used extensively in this work \citep{Jones_2001,Hunter_2007,Perez_2007,Astropy}.




\bibliography{../MyBibliography} 
\bibliographystyle{mnras} 




%
%


\bsp	
\label{lastpage}
\end{document}